\documentclass[aps,prd,twocolumn,notitlepage,preprintnumbers,nofootinbib,longbibliography]{revtex4-1}
\usepackage[utf8]{inputenc}
\usepackage{amsmath}
\usepackage{amssymb}
\usepackage{color}
\usepackage{braket}
\usepackage{slashed}
\usepackage{xcolor}
\usepackage{graphicx}
\usepackage{mathrsfs}
\usepackage{subfig}
\usepackage{tikz}
\usepackage[pdfpagelabels=false,colorlinks=true,urlcolor=teal,citecolor=teal]{hyperref}

\newcommand{\rd}{\mathrm{d}}

\newcommand{\nn}{\nonumber}

\newcommand{\Ecm}{{E_\text{cm}}}
\newcommand{\e}{\epsilon}
\newcommand{\rT}{\mathcal{T}}

\newcommand{\cB}{\mathcal{B}}

\newcommand{\cO}{\mathcal{O}}
\newcommand{\cC}{\mathcal{C}}

\newcommand{\cP}{{\mathcal P}}

\definecolor{darkred}{rgb}{0.7,0.0,0.0}
\definecolor{darkblue}{rgb}{0.0,0.0,0.9}
\definecolor{darkgreen}{rgb}{0.0,0.5,0.0}
\definecolor{brown}{rgb}{0.0,0.0,0.0}
\newcommand{\red}{\color{darkred}}
\newcommand{\blue}{\color{darkblue}}
\newcommand{\green}{\color{darkgreen}}

\newcommand{\cyan}{\green}

\def\ib{{\; \bar\imath}}

\begin{document}
\preprint{
MPP-2025-162
}
\preprint{
CPTNP-2025-047
}
\title{Energy Correlators Resolving Proton Spin}

\author{Jun Gao}
\email{jung49@sjtu.edu.cn}
\affiliation{State Key Laboratory of Dark Matter Physics, Shanghai Key Laboratory for Particle Physics and Cosmology, Key Laboratory for Particle Astrophysics and Cosmology (MOE),
School of Physics and Astronomy, Shanghai Jiao Tong University, Shanghai 200240, China}

\author{Hai Tao Li}
\email{haitao.li@sdu.edu.cn}
\affiliation{School of Physics, Shandong University, Jinan, Shandong 250100, China}

\author{Yu Jiao Zhu}
\email{yzhu@mpp.mpg.de}
\affiliation{Max-Planck-Institut f\"{u}r Physik, Werner-Heisenberg-Institut, Boltzmannstrasse 8, 85748 Garching, Germany}

\begin{abstract}
We investigate the partonic origin of the proton longitudinal spin using spin-dependent energy correlators  measured in  lepton-hadron collisions with longitudinally polarized proton beams. These observables encode angular correlations in energy flow and are sensitive to the spin-momentum structure of confined partons. Using soft-collinear effective theory, we analyze the correlation patterns in both nearly back-to-back and forward limits, which establishes a direct correspondence with longitudinally polarized transverse momentum-dependent distributions (TMDs) and nucleon energy correlators (NECs).  The TMDs and NECs allow consistent matching onto hard radiation regions and provide a comprehensive description of the transition from perturbative parton branching to nonperturbative confinement.
Using renormalization-group evolution, we obtain quantitative predictions
for spin-dependent energy correlation patterns at
next-to-next-to-next-to-leading logarithmic (N$^3$LL) accuracy in the
current fragmentation region and next-to-next-to-leading logarithmic
(NNLL) accuracy in the target fragmentation region.
 The framework provides new theoretical insight into how the internal motion and spin of partons contribute to the formation of the proton longitudinal spin and offers an experimental paradigm for probing the interplay between color confinement and spin dynamics at the forthcoming  Electron-Ion Collider.
\end{abstract}
\maketitle

\section{Introduction}
The origin of the proton spin has long been a central puzzle in QCD. 
In polarized deep inelastic scattering (DIS), the key observable is the first moment of the spin-dependent structure function
\begin{align}
\Gamma_1^P(Q^2) \;=\; \int_0^1 g_1(x,Q^2)\,\rd x .
\end{align}
Ellis and Jaffe, assuming vanishing strange-quark polarization, predicted a large positive value, 
$\Gamma_1^P \approx 0.186$~\cite{Ellis:1973kp}. 
The European Muon Collaboration measurement in 1988~\cite{EuropeanMuon:1987isl,EuropeanMuon:1989yki}, 
however, reported a much smaller value, $\Gamma_1^P \approx 0.126$, 
in striking disagreement with the prediction, thereby igniting the ``proton spin crisis".

To address this problem,  major progress has been made on both theoretical and experimental fronts, see Ref.~\cite{Ji:2020ena} for an overview.
On the experimental side, semi-inclusive DIS (SIDIS) has provided access to the flavor dependence of polarized quark distributions~\cite{HERMES:2004zsh,COMPASS:2009kiy}, while polarized $pp$ scattering at the Relativistic Heavy Ion Collider has constrained gluon and sea-quark helicities~\cite{RHICSPIN:2023zxx,Aidala:2012mv,Deur:2018roz}. These data have been incorporated into recent global analyses of polarized parton distribution functions  (PDFs)~\cite{Borsa:2024mss,Bertone:2024taw,Cruz-Martinez:2025ahf}. Looking ahead, the Electron-Ion Collider (EIC)~\cite{AbdulKhalek:2021gbh,AbdulKhalek:2022hcn} will provide unprecedented sensitivity to the quantitative determination of the partonic contributions to the proton spin.

On the theory side, two complementary decompositions of the proton spin have been established: the frame-independent ``Ji sum rule''~\cite{Ji:1996ek,Ji:1997pf}  and the partonic Jaffe-Manohar sum rule~\cite{Jaffe:1989jz} in the infinite momentum frame 
\begin{align}
    \frac{1}{2} = \tfrac12 \Delta\Sigma + \Delta G + \ell_q + \ell_g\,,
\end{align}
where the first two terms are the first moments of quark and gluon helicity distributions,
whose scale evolution is known to next-to-next-to-leading order (NNLO) accuracy~\cite{Moch:2014sna,Moch:2015usa,Blumlein:2022gpp,Blumlein:2021enk,Blumlein:2021ryt,Zhu:2025gts}. 
The last two terms, $\ell_q$ and $\ell_g$, 
denote higher-twist quark and gluon orbital angular momentum contributions.
For example, it has been proposed in quark models that $\ell_q$ is related to the  transverse momentum-dependent (TMD) pretzelosity distribution $h_{1T}^{\perp q}$~\cite{She:2009jq,Avakian:2009jt,Avakian:2010br}.

Significant progress has been made in addressing the proton spin puzzle from the lattice QCD perspective. 
The theoretical foundation was established in Ref.~\cite{Ji:1996ek} through  gauge-invariant decomposition of the nucleon spin, subsequent developments~\cite{Ji:2014lra,Zhao:2015kca} have made it possible to connect this framework with lattice simulations, 
and recent studies~\cite{Alexandrou:2020sml,Alexandrou:2017oeh,Yang:2016plb} have reported direct numerical results for the quark and gluon contributions to the proton spin.

On the small-$x$ front, helicity evolution equations derived by Kovchegov-Pitonyak-Sievert resum the single and double logarithms, 
leading to quantitative predictions for the small-$x$ asymptotics of the quark/gluon helicity distributions and the potential spin carried at very low $x$~\cite{Kovchegov:2016weo,Kovchegov:2015pbl,Kovchegov:2021lvz,Cougoulic:2022gbk,Kovchegov:2021lvz}.

Perturbative QCD, in addressing the proton spin puzzle through factorization and global analyses that solve the  ``inverse problem'' of determining  uncountably many low-energy constants, has now reached the NNLO frontier with the inclusion of higher-order radiative corrections~\cite{Ravindran:2003gi,Boughezal:2021wjw,Borsa:2020ulb,Borsa:2020yxh,Borsa:2022irn,Borsa:2022cap,Blumlein:2022gpp}.
While these approaches rely primarily on polarized DIS and $pp$ scattering cross sections, it is natural to seek complementary observables offering additional insight into the spin structure.
Energy-energy correlators (EECs), 
originally proposed in~\cite{Basham:1977iq,Basham:1978bw,Basham:1978zq,Basham:1979gh} as a means to investigate QCD dynamics, 
provide precisely such an opportunity.
Over the years, EECs have emerged as a versatile tool for exploring QCD dynamics across different regimes. 
In particular, they probe TMD dynamics through back-to-back jets~\cite{Collins:1981zc,deFlorian:2004mp,Moult:2018jzp,Duhr:2022yyp,Aglietti:2024xwv,Korchemsky:2019nzm,Chen:2023wah,Moult:2019vou,Kardos:2018kqj,Tulipant:2017ybb,vonKuk:2024uxe,Aglietti:2024zhg,Kang:2024dja,Gao:2019ojf,Gao:2023ivm} and reveal detailed jet substructure information in the collinear limit~\cite{Lee:2024icn,Lee:2024tzc,Liu:2024lxy,Dokshitzer:1999sh,Lee:2024esz,Chen:2024nyc,Konishi:1978ax,Konishi:1979cb,Konishi:1978yx,Dixon:2019uzg,Chen:2019bpb,Chen:2023zzh,Craft:2022kdo,Chen:2021gdk,Chen:2020adz,Chen:2025rjc,Gao:2025evv,Chen:2023zlx,Jaarsma:2023ell,Budhraja:2024tev,Song:2025bdj}.  EECs have been generalized to the DIS framework~\cite{Li:2020bub,Ali:2020ksn,Li:2021txc,Cao:2023qat,Kang:2023big}. As collinear-safe observables, they provide a robust probe of QCD dynamics at both perturbative and nonperturbative levels. Their versatility makes them a promising tool for testing the Standard Model with high precision, constraining parton distribution functions, and studying QCD factorization in new kinematic regimes. A comprehensive review can be found in Ref.~\cite{Moult:2025nhu}.  

In this work, we extend the EEC formalism to the case of a polarized proton target in DIS, defined as  follows
\begin{align}
 \langle \mathcal{E}(\hat n)  \rangle_{N}=\int \rd^4 x\, e^{i q\cdot x}
\langle P_N, S_{\parallel} | J(x)  {\mathcal{E}}(\hat n)
J(0) | P_N, S_{\parallel} \rangle\,,
\end{align}
where $N$ denotes the proton target and $\hat n$ is the generic radial unit vector where the energy weight is applied.  
Compared to conventional $q_T$-based observables, energy correlators offer several distinct advantages. First, owing to their strict operator definition, they provide a promising way to probe the nonperturbative spin structure of the proton from first principles. Second, as collinear-safe observables, EECs can be consistently defined across the full kinematic regime, from the hard radiation region to the current and target fragmentation regions. This seamless definition enables consistent matching across the dynamical domains, thereby offering a unique opportunity to investigate the transition from perturbative parton branching to nonperturbative confinement.

The remainder of this paper is organized as follows. In Sec.~\ref{sec:definition}, we review the definition of Bjorken-$x$ weighted energy correlators (ECs). Section~\ref{sec:current} discusses the factorization and resummation in the current region, while Sec.~\ref{sec:target} presents the corresponding analysis in the target region. 
The matching of polarized  nucleon energy correlators (NECs) to  polarized PDFs is carried out in Sec.~\ref{sec:matching}. Finally, we summarize our findings in Sec.~\ref{sec:summary}.
\section{Constructing EC in Deep Inelastic Scattering}
\label{sec:definition}
In this paper, we consider the polar angular $\theta$ distribution  in lepton-hadron colliders 
with a polarized high-energy lepton beam scattering off a longitudinally polarized nucleon target:
\begin{align}
 \ell(p_\ell,\lambda_\ell) + N(P_N, S_{\parallel}) \to \ell'(p_{\ell'},\lambda_{\ell'})  + X(P_X)
\,,\end{align}
where $X(P_X)$ denotes all the possible hadrons in the final states.
We will parametrize the events by the angular variable $\theta$, defined to be the polar angle between the detected particles (jets) with respect to proton velocity.
The leptonic momentum transfer and the invariant mass of the virtual photon are given by  $q=p_l-p_{\ell'}$ and $Q^2=-q^2$, respectively.
Additionally, $\lambda_\ell$/$S_{\parallel}$  denotes the chirality of the incoming lepton/target. 
The standard set of kinematic invariants is given by
\begin{align}
x=\frac{Q^2}{2 P_N\cdot q}\,\quad y=\frac{P_N\cdot q}{P_N\cdot p_\ell}\,,\quad z=\frac{P_N\cdot P_h}{P_N\cdot q}\,.
\end{align} 
The energy correlator is defined in the Breit frame, where the  proton carries the momentum
\begin{align}
P_N=\frac{Q}{2 x}=\frac{Q}{2 x} (1\,,0\,,0\,,1)\,,
\end{align}
and  the virtual photon  carries  momentum in its $z$ component
\begin{align}
q=\frac{Q}{2}(\bar n -n)=Q (0\,,0\,,0\,,-1)\,.
\end{align}
We are interested in energy-weighted angular $\theta$ correlations 
\begin{align}
\label{eq:eec-def}
\rd \sigma^{\text{EC}}
=&
\frac{16 \pi^2 \alpha^2}{2 E^2_\text{cm} Q^4}
\frac{\rd^3 p_{\ell'}}{(2\pi)^3 2E_{\ell'}} (4\pi) L_{\mu\nu}(p_\ell,p_{\ell'})W_{\text{EC}}^{\mu\nu}(q,P_N,\theta)
\nn\\
=&
\frac{2\pi\alpha^2}{Q^4}y \frac{\rd Q^2}{E^2_\text{cm}}\frac{\rd x}{x} 
L_{\mu\nu}(p_\ell,p_{\ell'})W_{\text{EC}}^{\mu\nu}(q,P_N,\theta)\,,
\end{align}
where $\alpha=e_R^2/(4 \pi)$ denotes the electromagnetic coupling, and instead of the conventional DIS hadronic tensor, the energy-flow operator is inserted between the electromagnetic currents
 \begin{align}
W_{\text{EC}}^{\mu\nu}&(q,P_N,\theta)
=\prod_X \frac{\rd^3 P_X}{(2\pi)^3 2 E_X}(2\pi)^4 \delta^{(4)}(q+P_N-P_X) 
\nn\\
\,\,\times&\frac{1}{4\pi}\langle P_N, S_{\parallel} | J^{\dagger \mu}  \hat{\mathcal{E}}(\theta)|X\rangle  
\langle X | J^\nu | P_N, S_{\parallel} \rangle \,,
\nn\\
=&
 \frac{1}{4\pi}\int \rd^4 x\, e^{i q\cdot x}
\langle P_N, S_{\parallel} | J^{\dagger \mu}(x)  \hat{\mathcal{E}}(\theta)
J^\nu | P_N, S_{\parallel} \rangle
\,.
\end{align}
The cumulant energy correlation operator quantifies the energy deposited in the detector for radiation within a radial angle bounded above by $\theta$,
\begin{align}
\hat{\mathcal{E}}(\theta)|X\rangle = \sum_{h \in X}\frac{E_h}{E_N}\Theta(\theta-\theta_h)|X\rangle\,.
\end{align}
From Poincaré covariance and parity symmetry,
the energy correlation hadronic tensor allows decomposition into independent hadronic structure functions~\cite{Collins:2011zzd}
\begin{align}
W_{\text{EC}}^{\mu\nu}(q,P_N,\theta)=f_T^{\mu \nu} x W_T  + f_L^{\mu \nu} x W_L+ \Delta f^{\mu \nu} x g_1\,, 
\end{align}
where the tensor bases are explicitly given in Appendix~\ref{sec: form-factor}.
We adopt the commonly used notation in which $W_T$ denotes the contribution from transverse modes of the virtual photon, while $W_L$ represents the longitudinal contribution.
The polarized hadronic function $g_1$ contributes to single spin asymmetry $1/2 (\sigma(S_{\parallel}=+)-\sigma(S_{\parallel}=-))$.
 Combining  the  hadronic tensor with
 the lepton current
\begin{align}
L^{\mu\nu}=2 \delta_{\lambda_\ell \lambda_{\ell'}}\left[ \left( p_\ell^\mu p_{\ell'}^\nu+p_\ell^\nu p_{\ell'}^\mu -\frac{Q^2}{2} g^{\mu\nu}\right)+i \lambda_l \epsilon^{\mu\nu \ell \ell'} \right]\,,
\end{align}
 we have decomposition as follows
  \begin{align}
  \label{eq:hadronic-scalar}
 \frac{\rd \sigma^{\text{EC}}}{\rd Q^2}=\delta_{\lambda_\ell \lambda_{\ell'}}
 \frac{2\pi\alpha^2}{Q^4}\rd x
 \left(
 \sigma_T W_T +\sigma_L W_L + \lambda_{\ell} S_{\parallel} \, \Delta\sigma\,  g_1
 \right)\,,
 \end{align} 
where the Born-level form factors are given by
\begin{align}
\label{eq:DIS-form}
\sigma_T=1+(1-y)^2\,,\quad \sigma_L=2-2y\,, \quad \Delta \sigma = 1-(1-y)^2\,.
\end{align}
We will also investigate Bjorken-$x$ weighted ECs, 
defined as the Mellin transform of the $x$ space ECs
\begin{align}
\label{eq:mellin-EEC}
\frac{\rd \sigma^\mathrm{EC}(N,\theta)}{\rd Q^2} 
= &\sum_h \int  x^{N-1}\rd \sigma(x,Q, P_h) \frac{E_h}{E_N}\Theta(\theta-\theta_h)
\nn\\
= &\int  x^{N-1} \frac{\rd \sigma^{\text{EC}}(x,Q, \theta)}{\rd Q^2}\,.
\end{align}
The DIS energy correlation exhibits remarkable simplicity in specific kinematic limits, offering novel insights into the mechanisms of color confinement and origin of nucleon spin. For instance,  the small-$x$ limit with $x \to 0$  should probe Regge asymptotics of QCD governed by
Balitsky-Fadin-Kuraev-Lipatov  evolution~\cite{Balitsky:1978ic,Kuraev:1977fs} and beyond. The study presented here is devoted to two complementary angular limits: the back-to-back limit $\theta \to \pi$ and the collinear limit $\theta \to 0$, where we establish  all-order factorization theorems in terms of universal nonperturbative objects. Reliable theoretical predictions are then obtained through renormalization group evolution.   

\section{Current Fragmentation and TMD Factorization}
\label{sec:current}
\begin{figure}
\centering
\includegraphics[width=0.35 \textwidth]{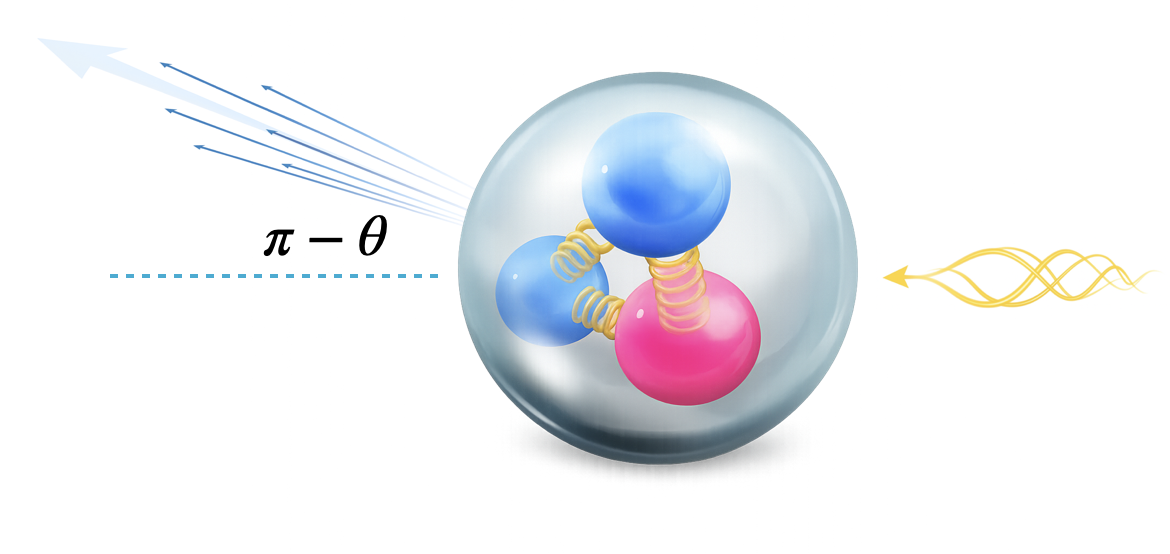}
\caption{Current fragmentation.}
\label{fig:EEC_Pi}
\end{figure}
In this section, we are concerned with back-to-back limit $\theta\to\pi$ where the outgoing quark
fragments into QCD jets, see Fig.~\ref{fig:EEC_Pi}. This kinematic region is also referred to as  current fragmentation region (CFR),
and the appropriate theoretical framework for describing it is TMD factorization.
To derive the factorization formulas, it is useful to identify a hadron inside the leading jet in the first place.
To this end, we consider  differential cross section with one identified hadron in the final states
\begin{align}
\frac{\rd \sigma_{\ell+N\to \ell'+h+X}}{\rd \vec q_\perp \rd x\, y\rd y\,  z\rd z   }=\frac{2 \pi \alpha^2}{4Q^4} L_{\mu \nu}(P_\ell,P_{\ell'}) W^{\mu \nu}(q, P_N, P_h)\,,
\end{align}
where $\vec q_\perp$ is the transverse momentum of the virtual photon in the hadron-hadron frame
or is a parametrization of the transverse momentum of the detected  leading jet $\vec q_\perp=-\vec P_h^\perp/z$ in the photon-hadron frame~\cite{Collins:2011zzd}.
The hadronic tensor with $h$ being resolved is given by
\begin{align}
W^{\mu\nu}(q,P_N,P_h)=&\prod_X \frac{\rd^3 P_X}{(2\pi)^3 2 E_X} \delta^{(4)}(q+P_N-P_h
\nn\\
-&P_X) \times\langle S_{\parallel}, P_N| J^{\dagger \mu}|h,X\rangle  \langle h,X | J^\nu | N \rangle \,.
\end{align}
Kinematic constraints in  CFR require that all final-state radiation in $X$ be collimated, with momenta either collinear to  the incoming target $N$ or the observed hadron $h$, and the hadronic amplitude is factorized through~\cite{Feige:2014wja}
\begin{align}
\langle {\green{h}};    {X_{{\blue{N}}\green{h}}};X_{\red{ \text {s}}}|J^\mu|{\blue{N}}\rangle\simeq &
\,\mathcal{C}(Q^2,\mu)
 \langle X_{\red{ \text {s}}}| \rT \left[
  Y^\dagger_{\green{\bar n}}Y_{{\blue{n}}}
  \right]
  |0\rangle
  \nn\\
  \times &
\langle {\green{h}}, {X_{\green{h}}}|\bar\chi_{\green{{\bar n}}}|0\rangle
\gamma^\mu
\langle {X_{\blue{N}}}| \chi_{\blue{n}}| {\blue{N}}\rangle\,,
\end{align}
where $X_{{\blue{N}}\green{h}}={X_{\blue{N}}}\otimes X_{\green{h}}$  denotes the collinear states and $X_{\red{ \text {s}}}$ denotes the soft states emitted from  the Wilson lines $Y^\dagger_{\green{\bar n}}$ or $Y_{{\blue{n}}}$~\cite{Feige:2014wja}.
We have absorbed the soft zero-bin subtraction into the normalization of the collinear fields,
since the matrix element computed with zero-bin subtracted soft-collinear eﬀective theory (SCET)  Lagrangian is equivalent
to dividing the naive matrix element by vacuum expectation values of zero-bin Wilson lines~\cite{Feige:2014wja,Lee:2006nr}.
The hard amplitude $\mathcal{C}(Q^2,\mu)$ is  insensitive to the  infrared (IR) dynamics of   soft and collinear degrees of freedom,
and thus can be computed at the partonic level. 
It is advantageous to consider $q+\gamma^\ast \to q$, with no soft radiation present
\begin{align}
\label{eq:scet-qqb}
\langle {\green{q}}|J^\mu| {\blue{q}}\rangle = 
\,\mathcal{C}(Q^2,\mu)
\langle {\green{q}}|\bar\chi_{\green{{\bar n}}}|0\rangle
\gamma^\mu
\langle 0| \chi_{\blue{n}}| {\blue{q}}\rangle\,.
\end{align}
In this case, the leading-power collinear expansion is exact ($=$ instead of $\simeq$), 
as each collinear trajectory contains just one particle---taking the collinear limit simply does nothing.
By construction, the IR divergences of the partonic amplitude $\langle {\green{q}}|J^\mu| {\blue{q}}\rangle$
is reproduced by the UV divergences of the hard scattering operators in SCET~\cite{Bauer:2000ew,Bauer:2000yr,Bauer:2001ct,Bauer:2001yt}.
Indeed,  the right-hand side of Eq.~(\ref{eq:scet-qqb}) is
\begin{align}
\label{eq:pole-transmu}
  \langle {\cyan  q}| \cO^{\vec \lambda}_\text{scet} |{\blue q} \rangle
 =  \langle {\cyan  q}| \cO^{\vec \lambda}_\text{scet} | {\blue q}\rangle_\text{tree}\left(1+\delta_{Z_{\vec \lambda}} (\epsilon_\text{IR})\right)\,,
\end{align}
where $\cO^{\vec \lambda}_\text{scet}\sim\bar\chi^\pm_{\green{{\bar n}}} \gamma^\mu \chi^\pm_{\blue{n}}\,,\,\,\, \lambda=\pm$\,.
The result is  proportional to tree-level `t Hooft--Veltman~\cite{tHooft:1972tcz}  bases because pure virtual-loop corrections in SCET
are scaleless and vanish identically. 
As a result, the IR poles of a  helicity amplitude is reproduced
 by UV renormalization counterterm diagrams provided by $\delta  {Z_{\vec \lambda}}= {Z_{\vec \lambda}}-1$~\cite{Moult:2015aoa}.

 In $q_T$ factorization, several subtleties need to be addressed. 
 The main issue concerns the validity of factorization itself, 
 namely demonstrating that Glauber effects drop out at the cross section level, 
 thereby justifying the separation of modes.
Once the hard- and soft-collinear degrees of freedom are disentangled, 
the next issue  is to consistently account for soft-collinear particles in a manner free of double counting.
Since these modes have the same invariant mass, 
they are distinguished by their rapidity, and the phase space integral must be regulated with a rapidity cutoff  $\nu$~\cite{Li:2016axz}.
Additionally,  a zero-bin subtraction is performed to remove the overlapping  between collinear and soft sectors,
yielding genuine collinear beam functions~\cite{Manohar:2006nz}. With these ingredients in place, we obtain the factorization  formula~\cite{Meng:1991da,Meng:1995yn,Nadolsky:1999kb,Ji:2004xq,Koike:2006fn,Collins:2011zzd}
\begin{align}
&W^{\mu\nu}\simeq
\frac{4}{z}
\sum_f  H_f(Q^2,\mu)
\int \frac{\rd^2  \vec b_\perp}{(2\pi)^2}
e^{-i \vec b_\perp\cdot \vec q_\perp} 
 \mathcal{S}_{n\bar n}(b_\perp,\mu,\nu)\times
 \nn\\
 &\mathrm{Tr}[
 \mathscr{B}_{f/N}(S_{\parallel},
 x,b_\perp,E_n,\mu,\nu)
\gamma^\mu
\mathscr{F}_{h/f}\left(z,\frac{b_\perp}{z},
E_{\bar n},\mu,\nu\right)
\gamma^\nu
]\,.
\end{align}
By applying  SCET Fierz identity 
\begin{align}
1 \otimes 1=\frac{1}{2} \biggl[ \frac{\slashed{\bar n}}{2} \otimes \frac{\slashed n}{2}-\frac{\slashed{\bar n} \gamma_5}{2} \otimes \frac{\slashed n \gamma_5}{2}-\frac{\slashed{\bar n} \gamma_\perp^\mu}{2} \otimes
 \frac{\slashed n \gamma_{\perp\mu}}{2} \biggr]\,,
\end{align}
 the  resulting SIDIS TMD factorization formula is
 \begin{widetext}
\begin{align}
\label{eq:SIDIS-fac}
x y z \frac{\rd \sigma_{\ell+N\to \ell'+h+X}}{\rd^2  \vec q_\perp \rd x\rd y\rd z}
\simeq&\, \sum_f  H_f(Q^2,\mu)  \int \frac{\rd^2  \vec b_\perp}{(2\pi)^2} e^{-i \vec b_\perp\cdot \vec q_\perp}\,
\mathcal{S}_{n\bar n}\left(b_\perp\,,\mu\,,\nu\right)
z\mathcal{F}_{h/f}\left(z\,,\frac{b_\perp}{z}\,,E_{\bar n}, \mu,\nu\right)
\nn\\
\times
&
\left(
\sigma^\text{U}_0\times x\mathcal{B}_{f/N} (x\,,b_\perp\,,E_n, \mu,\nu) 
+\lambda_\ell S_{\parallel}\times\sigma^\text{L}_0\times x \Delta \mathcal{B}_{f/N} (x\,,b_\perp\,,E_n, \mu,\nu)
\right)\,,
\end{align}
\end{widetext}
where  
$H_f(Q^2,\mu)=|\cC_f(Q,\mu)\rangle\langle \cC_f(Q,\mu)|$
represents the  square of the hard matching coefficient~\cite{Becher:2006mr,Moch:2005tm,Moch:2005id,Gehrmann:2010ue,Baikov:2009bg,Gehrmann:2010tu,Lee:2022nhh},
 and $E_n$ ($E_{\bar{n}}$) are the energies of  struck (fragmented)  partons.
The born-level unpolarized cross section $\sigma^\text{U}_0$ and its asymmetry counterpart $\sigma^\text{L}_0$ are 
\begin{align}
\sigma^\text{U}_0=2\pi \alpha^2 \frac{1+(1-y)^2}{Q^2}\,,\quad \sigma^\text{L}_0=2\pi \alpha^2 \frac{1-(1-y)^2}{Q^2} \,.
\end{align}
The TMD beam  and fragmentation functions are  the  genuine collinear contributions with zero-bin subtraction
\begin{align}
    \mathcal{B}_{q/N}& (x\,,b_\perp\,,E_n, \mu,\nu)
    \equiv
    \lim_{\nu\to\infty}
   \frac{\mathcal{B}^{0}_{q/N} (x\,,b_\perp\,,E_n, \mu,\nu)}{\mathcal{S}^{0}_{n\bar n}(b_\perp\,,\mu,\nu)}\,,
\end{align}
the superscript $0$ denotes  UV-renormalized but  unsubtracted collinear beam functions.
With an exponential rapidity regulator~\cite{Li:2016axz}, the  zero-bin soft function is identical to the  SIDIS TMD soft function, 
the latter is  related to back-to-back EEC soft function~\cite{Li:2016ctv} by a boost
\begin{align}
\mathcal{S}_{n\bar n}\left(b_\perp\,,\mu\,,\nu\right)=\mathcal{S}_\text{EEC}\left(b_\perp\,, L_{\nu}+\ln\frac{n \cdot  \bar n}{2}\,,   \mu \right)\,,
\end{align}
where $ L_\nu=\ln\left({\nu^2}/{\mu^2}
  \right)$.
The explicit expression of the TMD soft function can be found in Refs.~\cite{Li:2016ctv,Moult:2018jzp,Moult:2022xzt}.
In this work we're concerned with energy correlations between the proton and the backward leading jets. 
To this end, we have already
weighted SIDIS TMD factorization formula in Eq.~(\ref{eq:SIDIS-fac}) with energy weight $E_h/E_N\simeq x z$,
and, subsequently, summed over all possible hadrons 
\begin{align}
\frac{\rd \sigma^\mathrm{EC}_{\ell+N\to \ell'+X}}{\rd^{2} \vec q_\perp \rd x\rd y}=\sum_h \int \! \rd z \,\, x  z \frac{\rd \sigma_{\ell+N\to \ell'+h+X}}{\rd^2  \vec q_\perp \rd x\rd y\rd z}\,,
\end{align}
to obtain
\begin{widetext}
\begin{align}
\label{eq:fac-beam}
 y  \frac{\rd \sigma^\mathrm{EC}_{\ell+N\to \ell'+X}}{\rd^{2} \vec q_\perp \rd x\rd y}
\simeq&\,
\sum_f 
\int \frac{\rd^2 \vec b_\perp}{(2\pi)^2} e^{-i \vec b_\perp \cdot\vec q_\perp}\,
\left[H_f(Q^2,\mu) \mathcal{S}_{n\bar n}\left(b_\perp\,,\mu\,,\nu\right)\right]
\mathcal{J}_{f}\left(b_\perp, E_{\bar n}, \mu,\nu\right)
\nn\\
\times&
\left(
\sigma^\text{U}_0  \times  x\,\mathcal{B}_{f/N} (x\,,b_\perp\,,E_n, \mu,\nu) 
+
\lambda_\ell S_{\parallel}\times\sigma^\text{L}_0\times x\,\Delta\mathcal{B}_{f/N} (x\,,b_\perp\,,E_n, \mu,\nu) 
\right)
 \,,
\end{align}
\end{widetext} 
where we perform the twist-$2$ matching, namely the small-$b$ operator product expansion (OPE), 
of the TMD fragmentation function (FF) onto collinear FFs
\begin{align}
\mathcal{F}_{h/f}\left(z\,,\frac{b_\perp}{z}\,,E_{\bar n}, \mu,\nu\right)
=&\sum_i  d_{h/i} \otimes \mathcal{C}_{if}+ \mathcal{O}(b_T^2\Lambda^2_{\text{QCD}}) \,,
\end{align}
and  use FF sum rules to obtain the EEC TMD jet function $\mathcal{J}_f$~\cite{Moult:2018jzp},
\begin{align}
\mathcal{J}_f(b_\perp\,,E_{\bar n},\mu,\nu)
\equiv&
\sum_h \int_{0}^{1} \rd z\, z\, \mathcal{F}^\text{OPE}_{h/f}\left(z\,,\frac{b_\perp}{z}\,,E_{\bar n},\mu,\nu\right)
\nn\\
= &
\sum_i\int_{0}^{1}\rd \xi \,\xi\, \mathcal{C}_{if}\left( \xi ,\frac{b_\perp}{\xi} ,E_{\bar n},\mu,\nu\right)\,.
\end{align}
Nonperturbative corrections to this function have been recently analyzed in Refs.~\cite{Kang:2024dja,Cuerpo:2025zde}.

For the experimental extraction of the TMDs,
we express the factorization formula in terms of 
physical TMDs and jet functions,  which are obtained by multiplying the former by the square root of the TMD soft function
\begin{align}
f^q_1 \left(x\,,b_\perp\,,\xi^n, \mu\right)=&\,\mathcal{B}_{q/N} (x\,,b_\perp\,,E_n, \mu,\nu)\sqrt { \mathcal{S}_{n\bar n}}\,,
\nn\\
D^q_1\left(z\,,\frac{b_\perp}{z}\,,\xi^{\bar n}, \mu\right)=& \,\mathcal{F}_{h/q}\left(z\,,\frac{b_\perp}{z}\,,E_{\bar n}, \mu,\nu\right) \sqrt{\mathcal{S}_{n\bar n}}\,.
\end{align}
\begin{figure*}[htbp]
  \centering
    \subfloat[]{\includegraphics[width=0.28\textwidth]{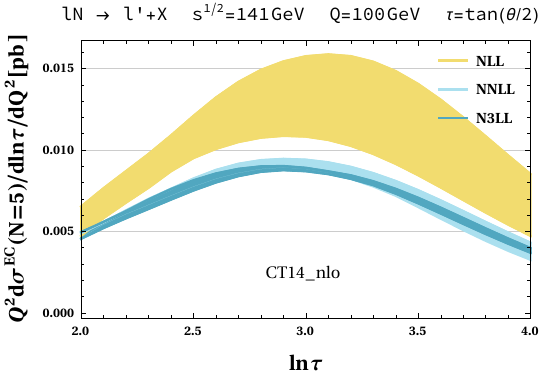}}\hfill
   \subfloat[]{\includegraphics[width=0.28\textwidth]{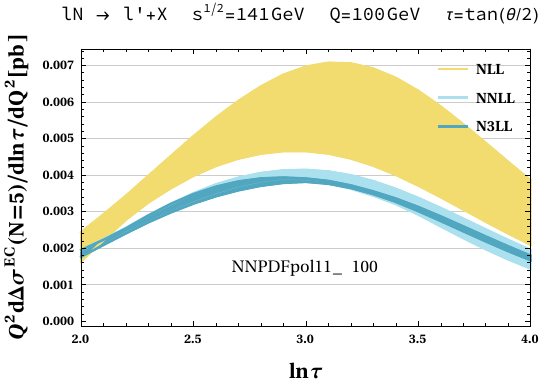}}\hfill
  \subfloat[]{\includegraphics[width=0.28\textwidth]{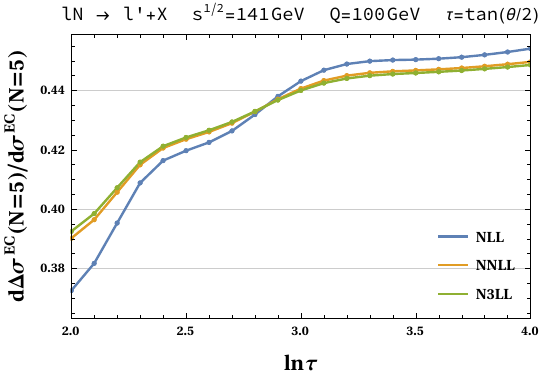}}
 \caption{ 
 (a), (b) Back-to-back limit of Bjorken-$x$ weighted ($N$=5) ECs using  BLNY18 nonperturbative model functions~\cite{Sun:2014dqm}.
(c) The ratio between polarized (b) and unpolarized (a) energy-weighted cross sections.
 }
\label{fig:EEC-btob}
\end{figure*}
As a result, the rapidity divergences $\ln(4 E_i^2/\nu^2)$ cancel between collinear and soft sectors, leaving  behind physical rapidity logarithms $\ln(\xi^i/\mu_b^2)$ with $\mu_b=b_0/b_T, b_0=2e^{-\gamma_E}$,
and $\xi^i$ are the Collins-Soper (CS) scales~\cite{Collins:1981uk,Collins:1987pm,Collins:2011zzd} for SIDIS 
\begin{align}
\label{eq:CS-scale}
\xi^n = Q^2 \frac{x}{x+y-x y}\,,\quad \xi^{\bar n} = Q^2 \frac{x+y-x y}{x}\,.
\end{align}
The $\mu$ evolution  of the hard matrix and the TMDs is linear in $\ln \mu$ to all-loop orders, as can be proved by  renormalization group (RG) consistency
\begin{align}
\label{eq:rg-CS-mu}
\frac{\rd}{\rd\ln\mu }\ln f_i (x_i, b_\perp, \xi^i,\mu) 
=\, - \gamma^i_{\rm cusp}\ln\frac{\xi^i}{\mu^2} -\gamma^i_{H} \,.
\end{align}
The rapidity evolution of the TMDs  is referred to as the  CS kernels 
\begin{align}
\label{eq:rg-CS-nu-1}
K_i(b_\perp,\mu)\equiv\frac{\rd}{\rd\ln \sqrt{ \xi^i}}\ln f_i (x_i, b_\perp, \xi^i,\mu)\,,
\end{align}
again by RG consistency, the $\mu$-evolution of $K_i(b_\perp,\mu)$ is controlled by the cusp anomalous dimension
\begin{align}
\frac{\rd K_i(b_\perp,\mu)}{\rd\ln\mu}  = -2 \gamma^i_{\rm cusp} (\alpha_s(\mu))\,.
\end{align}
From  above we can  solve the RG equation to give
\begin{align}
\label{eq:rg-CS-nu-2}
K_i(b_\perp,\mu) 
=&-2 A^i_{\rm cusp}(\mu,\mu_b)+\gamma^i_R(\mu_b)\,.
\end{align}
The solution has a evolution factor  $A^i_{\rm cusp}(\mu,\mu_b)$ controlled by the cusp anomalous dimension $\gamma^i_{\rm cusp}$;
it also leaves a residual boundary term $\gamma^i_R(\mu_b)$ known as QCD rapidity anomalous dimension.
While the cusp anomalous dimension $\gamma^i_{\rm cusp}$ is a purely perturbative object which originates from the UV renormalization of the cusped Wilson loops~\cite{Brandt:1981kf,Brandt:1982gz}, the rapidity anomalous dimension $\gamma^i_R(\mu_b)$, on the other hand,  has both perturbative part as $b_T \to 0$ and  genuine nonperturbative part as $b_T \to \infty$. The perturbative accuracy for the CS kernel has reached  up to N${}^4$LL~\cite{Li:2016ctv,Moult:2022xzt,Duhr:2022yyp},
and there has been tremendous progress toward lattice calculation for the nonperturbative part as well~\cite{LatticePartonLPC:2022eev,Shanahan:2020zxr,Shanahan:2021tst,Avkhadiev:2024mgd,Schlemmer:2021aij,Li:2021wvl,LatticeParton:2020uhz,Tan:2025ofx}.
Equations~(\ref{eq:rg-CS-mu})\,,\,(\ref{eq:rg-CS-nu-1})\,, and (\ref{eq:rg-CS-nu-2})
are referred to as
Collins-Soper equations~\cite{Collins:1981uk,Collins:1987pm,Collins:2011zzd},
they are fully equivalent to the modern language of SCET rapidity RGs~\cite{Chiu:2011qc,Chiu:2012ir} in the context of $q_T$ factorization.
Solving the CS equation, we will have the following RG-improved factorization using physical TMDs
\begin{widetext}
\begin{align}
\label{eq:SIDIS-fac-semi-CS}
y \frac{\rd \sigma^\mathrm{EC}_{\ell+N\to \ell'+X}}{\rd^{2} \vec q_\perp \rd x\rd y}
\simeq&\,
\sum_q 
\int \frac{\rd^2 \vec b_\perp}{(2\pi)^2} e^{-i \vec b_\perp \cdot\vec q_\perp}\,
H_q(Q^2,\mu)
\textrm{J}_q\left(\xi^{\bar n}_0,\mu^{\bar n}_0\right)
\times
\left(
\sigma^\text{U}_0\times x f_1^q(x,b_\perp,\xi^{n}_0,\mu^n_0)
+\lambda_\ell S_{\parallel}\,\sigma^\text{L}_0\times x g_1^q(x,b_\perp,\xi^{n}_0,\mu^n_0)
\right)
\nn\\
\times
&
\,\prod_ie^{-2 K^i_{\rm cusp}(\mu^i_0,\mu)+A^i_H(\mu^i_0,\mu)}
\times
\left(
\frac{\xi^i}{{\mu^i_0}^2}
\right)^
{
A^i_{\rm cusp}(\mu^i_0,\mu)
}
\times
\left(
\frac{\xi^i}{\xi^i_0}
\right)^
{
-A^i_{\rm cusp}(\mu^i_0,\mu_b)
+\frac{1}{2}
\gamma_R^i(\mu_b)
}
\,.
\end{align}
\end{widetext}
 \begin{figure}
\centering
\includegraphics[width=0.35 \textwidth]{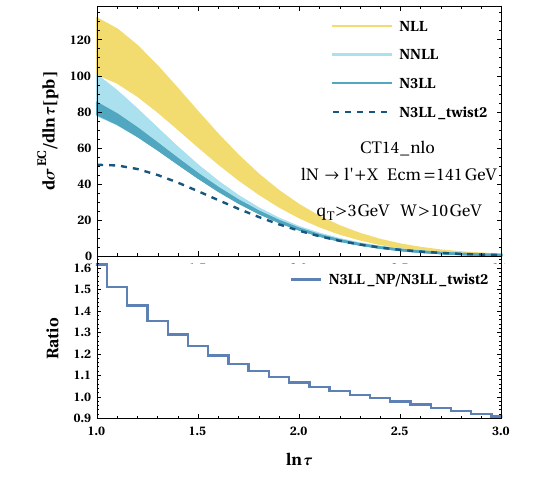}
\includegraphics[width=0.35 \textwidth]{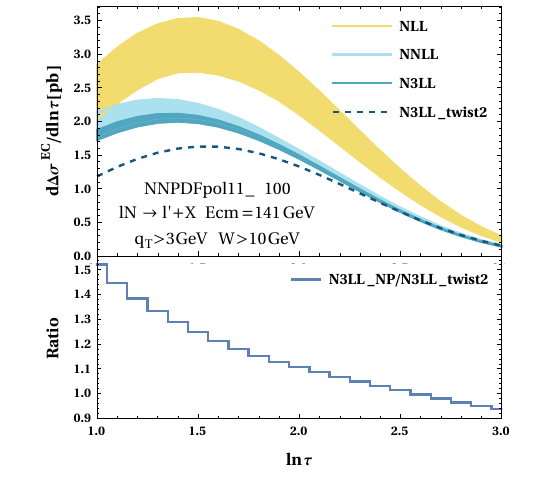}
\caption{Back-to-back limit of Bjorken $x$ weighted ($N$=1) ECs.
A transverse-momentum cut on $q_T$ is imposed to veto contributions from nonperturbative modes, 
and a hadronic invariant-mass cut $W=\sqrt{(P+q)^2}>10$GeV is applied to ensure a hard virtual-photon--nucleon collision.
The solid line (with scale uncertainty) is obtained  by implementing the $b_\ast$
 prescription together with the BLNY18 nonperturbative model functions~\cite{Sun:2014dqm}, 
 while the dashed line is obtained without applying either the $b_\ast$
 prescription or any nonperturbative model.
The ratio is obtained from a comparison of the central value of the solid line and dashed line at N${}^3$LL level.
}
\label{fig:cut_qt}
\end{figure}
Equations~(\ref{eq:fac-beam}) and (\ref{eq:SIDIS-fac-semi-CS}) capture the back-to-back limit of the 
energy correlations observable defined in Eq.~(\ref{eq:mellin-EEC}), upon performing an appropriate change of variables
from $(x,y,q_T)$ to $(x,Q,\ln \tau)$.
The relation between the jet transverse momentum $q_T$ and the hadron rapidity $Y$
is given by 
\begin{align}
\label{eq:qt-th}
\sin\theta=\frac{q_T}{Q/2}\,,\quad Y=-\ln\tau=-\ln \tan(\theta/2)\,,
\end{align}
while $y$ is related to the hard scale $Q$ through $Q^2=x y E^2_\mathrm{cm}$.
The Mellin-$N$ space results are obtained by Mellin transforming the $x$ space expressions.

Figure~\ref{fig:EEC-btob} displays the N${}^3$LL-resummed Bjorken-$x$ weighted energy correlators for both unpolarized and polarized cases.
In Fig.~\ref{fig:EEC-btob}(a) and \ref{fig:EEC-btob}(b), we present Mellin space results by weighting Eq.~(\ref{eq:SIDIS-fac-semi-CS}) with Bjorken weight $x^{N-1}$, we choose $N=5$ to suppress the small-$x$ contributions, and we have implemented BLNY18~\cite{Sun:2014dqm} nonperturbative model therein. 
Specifically, we approximate the initial-state TMDs by their twist-2 matchings onto collinear PDFs~\cite{Luo:2019hmp,Luo:2019bmw,Luo:2020epw,Ebert:2020qef,Ebert:2020yqt,Zhu:2025gts} 
while the higher-twist contributions are modeled  using the BLNY18 parametrizations given below
\begin{align}
\label{eq:syy2}
S_{\text{NP}}^{\text{BEAM}}=
{g_1 \over 2 }
 b^2
+ \frac{1}{2}g_2 \ln\left(\frac{b}{b_\ast}\right)\ln\left(\frac{Q}{Q_0}\right)
\,,
\end{align}
together with the  $b^\ast$ prescription~\cite{Collins:2014jpa} to avoid the Landau poles
\begin{align}
\label{eq:bstar}
b\to b^\ast=\frac{b}{\sqrt{1+b^2/b^2_\text{max}}}\,,\quad b_\text{max}=1.5\text{GeV}^{-1}\,.
\end{align}
We also quote here the parameter space in the model
\begin{align}
\label{eq:blny}
g_1=0.212\,,\quad g_2=0.84\,,\quad Q_0=1.55\text{GeV}\,.
\end{align}
For the nonperturbative corrections to the final-state EEC jet function, we follow the prescription  of Ref.~\cite{Li:2021txc}.
Reference \cite{Kang:2024dja} uses slightly different $b_\ast$ prescriptions, but the resulting numerical differences are small.
In Fig.~\ref{fig:EEC-btob} (c), we show the ratio of polarized and unpolarized energy-weighted cross sections.

For energy correlations in the back-to-back limit, only the collinear twist-2 contributions are known~\cite{Luo:2019hmp,Luo:2019bmw,Luo:2020epw,Ebert:2020qef,Ebert:2020yqt,Zhu:2025gts}, 
within uncertainties that reflect the propagation of collinear PDF uncertainties into the global fits of TMD distributions~\cite{Bury:2022czx}.
We use throughout this paper the \textsc{LHAPDF6}~\cite{Buckley:2014ana} PDF sets \textsc{CT14nlo}~\cite{Dulat:2015mca} and \textsc{NNPDFpol1.1}~\cite{Nocera:2014gqa} as the collinear twist-2 inputs.

The remaining dominant sources of uncertainty and model dependence arise from nonperturbative higher-twist effects of order $\mathcal{O}(b_T\Lambda_\mathrm{QCD})$,
which become relevant at large $b_T$ (or, equivalently when $q_T$ is of order $\Lambda_\mathrm{QCD}$),
as well as from subleading-power corrections in $\theta\sim\mathcal{O}(q_T/(Q/2))$,
corresponding to deviations of the observation angle from $\pi$.
In the BLNY18 fit, the twist-2 singular contribution is treated at next-to-leading logarithmic (NLL) accuracy. 
However, this level of accuracy is not well justified in light of the large radiative corrections observed when going from NLL to N${}^3$LL accuracy. 
On the other hand, the nonsingular contribution (referred to as the $Y$-term therein) is not systematically included in their analysis due to the lack of theoretical control over power corrections. 
It therefore remains an open question over which kinematic range the parameters in Eq.~(\ref{eq:blny}) can be reliably applied.

To this end, we consider energy correlations with a $q_T$ cut and choose $q_T^\text{cut} = 3\text{GeV}$.
As shown in Fig.~\ref{fig:cut_qt}, the result is stable under variations of $(\mu\,,\sqrt\xi)$ and of the upper limit $b_T^\mathrm{max}$
of the $b_T$ integral in Eq.~(\ref{eq:SIDIS-fac-semi-CS}), which is varied between 3  and 10GeV${}^{-1}$. 
This stability indicates that contributions from modes with $b_T>3$GeV${}^{-1}$ are effectively suppressed by the $q_T$ cut.
As a result, the observable is dominated by collinear twist-2 contributions, and the
 $b_\ast$ prescription can be dropped, the corresponding result is shown as the dashed line in Fig.~\ref{fig:cut_qt}.
From the figure, we observe that the solid and dashed curves coincide for $\ln \tau > 2$ (corresponding to angles larger than approximately 165$^\circ$), 
where the dynamics are dominated by perturbative resummation effects. In contrast, 
in the range $1 < \ln \tau < 2$ (corresponding to angles between roughly 140$^\circ$ and 165$^\circ$), the BLNY18 nonperturbative model introduces additional modifications. 
This intermediate region corresponds to the transition from the perturbative resummation regime to the hard-radiation-dominated region. 

These observations suggest that the BLNY18 fit requires further improvement, potentially through the inclusion of higher-order radiation corrections, 
both in the Sudakov region (extending from NLL to N${}^3$LL accuracy) and in the transition region (the  $Y$-term).

\section{Target Fragmentation  and  Nucleon Energy Correlators}
\label{sec:target}
\begin{figure}
\centering
\includegraphics[width=0.35 \textwidth]{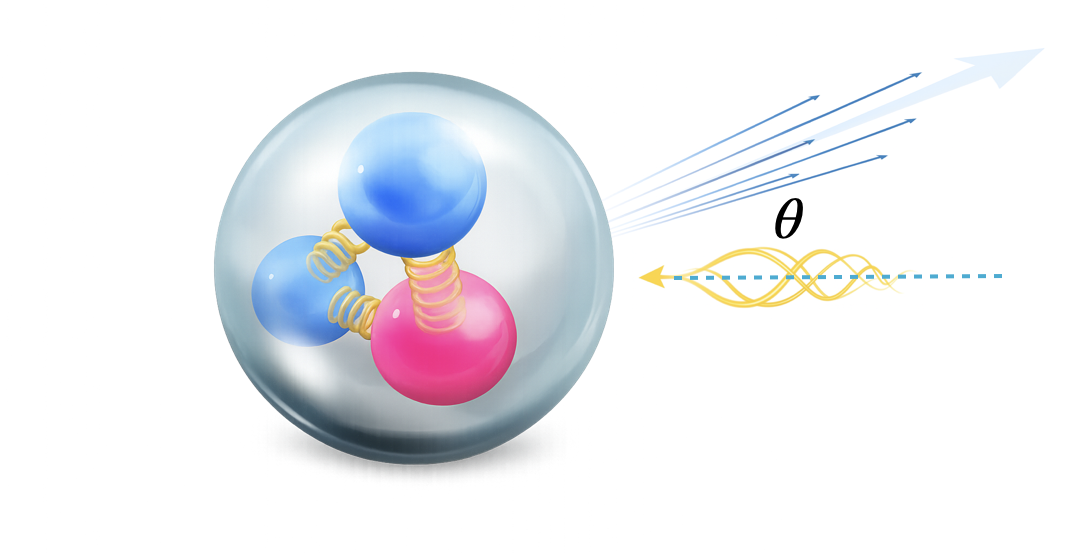}
\caption{Target fragmentation.}
\label{fig:EEC_0}
\end{figure}

In this section, we are concerned with collinear limit $\theta\to 0$ where the spectator partons
fragment into forward QCD jets, see Fig.~\ref{fig:EEC_0}. This kinematic region is also referred to as  the  target fragmentation region (TFR),
and the appropriate theoretical framework for describing it is through fracture functions~\cite{Trentadue:1993ka,Berera:1995fj,Grazzini:1997ih,deFlorian:1997wi,Ceccopieri:2007th,Anselmino:2011ss}.
Fracture functions provide a joint description of the struck parton distribution and the fragmentation of spectator partons into an identified hadron.
In Refs~\cite{Liu:2022wop,Cao:2023oef}, the NEC is proposed as an inclusive version of fracture functions  by integrating over the  momentum fraction of the identified hadrons and summing over all possible species of  them, the  sum rule is discussed in Ref~\cite{Chen:2024bpj}.
The NEC has been evaluated at NLL accuracy in Refs~\cite{Liu:2022wop,Cao:2023qat,Cao:2023oef}. 
In the  present work, we extend the calculation  to next-to-next-to-leading logarithmic (NNLL)  accuracy and, additionally, include proton longitudinal spin. 
We will first briefly review the factorization formulae to the  leading power of the limit $\theta\to 0$.
The complete set of  effective operators is obtained in Refs~\cite{Liu:2022wop,Cao:2023qat} by integrating out hard off-shell modes and is identified as  twist-2 operators with the insertion of energy-flow operators
\begin{align}
O_n^{i;\Gamma}(\omega_\pm,\theta)=&
    \frac{1}{\omega_+}
    \left[
    \bar \chi^{(i)}_{n,\omega_1}
    \,\Gamma\,
   \hat {\mathcal{E}}(\theta)
    \chi^{(i)}_{n,\omega_2}
 \right]\,,
 \nn\\
 O_n^{g;\Gamma}(\omega_\pm,\theta)=&
    \frac{-1}{\omega^2_+}
    \left[
    \cB^{\perp\mu}_{n,\omega_1}
   \, \Gamma_{\mu\nu} \,
      \hat{\mathcal{E}}(\theta)
        \cB^{\perp\nu}_{n,\omega_2}
 \right]\,,
\end{align}
where $\Gamma$s denote the quark/gluon spin projectors for an unpolarized/longitudinally polarized nucleon target
\begin{align}
\label{eq:spin-projectors}
    \Gamma \in \left\{    \frac{\slashed{\bar n} }{2}\,,
    \frac{\slashed{\bar n} \gamma_5}{2}
    \right\}\,,
    \quad
      \Gamma^{\mu\nu} \in \left\{   
      g_\perp^{\mu\nu}
      \,,
    i  \epsilon_\perp^{\mu\nu}
    \right\}\,,
\end{align}
and the gauge-invariant collinear building blocks with label energy $\omega$ ($\omega_\pm=\omega_1\pm \omega_2$) is defined by
$\chi_{n,\omega}
    \equiv 
\left[\delta(\omega-\bar \cP)\chi_n \right]$
    and
$ g\cB^{\perp\mu}_{n,\omega}
     \equiv  
        [g\cB^{\perp\mu}_n
   \delta(\omega-\bar \cP^\dagger) ]$.
 The hadronic scalar functions  in Eq.~(\ref{eq:hadronic-scalar})  
 are then matched onto the twist-2 operator bases~\cite{Bauer:2002nz}, for instance,
\begin{align}
\label{eq:t2-match}
  g_1(x,Q,\theta)\to & \sum_{i }  \int 
 \frac{\rd \omega_+ \rd \omega_-}{2}
  \,
    \mathscr{C}^{\Gamma}_{i}(\omega_\pm,x,Q,\mu)  O_n^{i;\Gamma}(\omega_\pm,\theta)\,,
\end{align}
where $\mathscr{C}^{\Gamma}_{i}$ are the matching coefficients of structure function $ g_1$,
the spin index $\Gamma$ should take the right-hand side of Eq.~(\ref{eq:spin-projectors})
as antisymmetric tensors.
To proceed, we use the relation between label operators and the conventional NECs, for example,
\begin{align}
\label{eq:label-pdf}
    \langle  P_N;S_\parallel|& \bar \chi_{n,\omega} \frac{\slashed{\bar n} \gamma_5}{2} 
     \hat {\mathcal{E}}(\theta)
    \chi_{n,\omega}|P_N;S_\parallel\rangle
    =
    4\bar n \cdot P_N S_\parallel
    \nn\\
    \times&
    \int_{-1}^1 \rd \xi \delta(\omega_-)
    \delta(\omega_+ - 2\xi\bar n\cdot P_N)\Delta F^\text{EC}_{i/N}(\xi,\theta)\,,
\end{align}
and charge conjugation symmetries of both the NECs
and  the Wilson coefficients 
\begin{equation}
     F^{\text{EC}}_{\ib/N}(\xi,\theta)= - F^{\text{EC}}_{i/N}(-\xi,\theta)\,,
     \Delta F^{\text{EC}}_{\ib/N}(\xi,\theta)=\Delta F^{\text{EC}}_{i/N}(-\xi,\theta)\,,
\end{equation}
to obtain the matching conditions for the unpolarized structure  function $W_T(x,Q,\theta)$, $W_L(x,Q,\theta)$ 
and the spin-asymmetry counterpart $g_1(x,Q,\theta)$.
For instance, the leading-power expansion   for the spin-asymmetry part is given by a convolution between perturbatively calculable
coefficient functions and the nonperturbative NECs
\begin{equation}
  g_1(x,Q,\theta)   =  \sum_{i=1}^{N_f}
        \mathscr{C}_{i}^\Gamma 
  \otimes
       (\Delta F^{\text{EC}}_{i/N}+ \Delta F^{\text{EC}}_{\ib/N})    
   +     
    \mathscr{C}_{g}^\Gamma
   \otimes
     \Delta F^{\text{EC}}_{g/N}   \,.
\end{equation}
It proves useful to rewrite the above factorization formula through singlet and nonsinglet combinations of the NECs,
for instance
\begin{align}
F^\text{EC}_{q/N}(\xi,\theta)=&\sum_{i=1}^{N_f} F^\text{EC}_{i/N}(\xi,\theta)+ F^\text{EC}_{\ib/N}(\xi,\theta)\,,
\nn\\
F^\text{EC,NS}_{i/N}(\xi,\theta) =&  F^\text{EC}_{i/N}(\xi,\theta)+ F^\text{EC}_{\ib/N}(\xi,\theta)-\frac{1}{N_f}F^\text{EC}_{q/N}(\xi,\theta)\,.
\end{align}
Finally, we gather  our factorization formulas   as follows
  \begin{align}
 \frac{\rd \sigma^{\text{EC}}}{\rd Q^2}=\delta_{\lambda_\ell \lambda_{\ell'}}
 \frac{2\pi\alpha^2}{Q^4}\rd x
 \left(
 \sigma_T W_T +\sigma_L W_L + \lambda_{\ell} S_{\parallel} \, \Delta\sigma\,  g_1
 \right)\,,
 \end{align} 
 where the scalar structure functions admit the following leading-power expansion as $\theta\to 0$
 \begin{widetext}
 \begin{align}
 \label{eq:fac-x-space}
 W_\Gamma (x,Q,\theta)\simeq &\,
\sum_i Q_i^2 \,\mathscr{C}_{i;\text{NS}}^\Gamma(Q,\mu) \otimes F^{\text{EC,NS}}_{i/N}(\theta,\mu) + \left( \frac{1}{N_f} \sum_{i=1}^{N_f} Q_i^2 \right) \, \bigg[
  \mathscr{C}_{q}^\Gamma(Q,\mu) \otimes F^\text{EC}_{q/N}(\theta,\mu) + \mathscr{C}_{g}^\Gamma(Q,\mu) \otimes F^\text{EC}_{g/N}(\theta,\mu)
 \bigg]
 \,.
 \end{align}
  \end{widetext}
The NEC provides a joint description of the struck parton distribution and the energy flows of forward QCD jets originating from spectator fragmentations.
The associated energy-flow operator $\hat{ \mathcal{E}} (\theta)$ acts only on the collinear sector,
thus, if  $\hat{ \mathcal{E}} (\theta)$ were set to the identity, i.e.\,, in the absence of detectors in the forward region, NECs reduce to standard PDFs and the resulting factorization formula coincides with that of inclusive DIS.
 From this perspective, the Wilson  coefficients $\mathscr{C}^{\Gamma}_{i}$  are identical to the DIS coefficient functions~\cite{Blumlein:2022gpp,
 Vermaseren:2005qc,Moch:2007rq,Moch:2008fj,Zijlstra:1993sh,Vogt:2008yw,Zijlstra:1992qd,Moch:1999eb}, 
 and the UV renormalization of NECs matches that of the conventional PDFs, for instance
\begin{align}
\label{eq:UV-renor}
  \begin{bmatrix}
F^{\text{EC}}_{q/N}\left(E_N \sin \theta,\mu\right) 
\\
 F^\text{EC}_{g/N}\left(E_N \sin \theta,\mu\right) 
  \end{bmatrix} 
   = \widehat{Z}_\text{S}^\text{PDF}
   \nn
   \otimes
     \begin{bmatrix}
F^{\text{EC}0}_{q/N}\left(E_N \sin \theta,\mu\right) 
\\
 F^{\text{EC}0}_{g/N}\left(E_N \sin \theta,\mu\right)
  \end{bmatrix}    \,,
\end{align}
where  the NECs  are and must be parametrized by hadronic logarithm $\ln E_N \sin \theta/\mu$,
with $E_N$ being the energy of the incoming target $N$, in DIS $E_N = Q/(2x)$.
The multiplicative renormalization factor $ \widehat{Z}_\text{S}^\text{PDF}$  
is the standard PDF UV renormalization factor 
for the singlets, and the associated 
Dokshitzer-Gribov-Lipatov-Altarelli-Parisi (DGLAP)~\cite{Gribov:1972ri,Lipatov:1974qm,Altarelli:1977zs} evolution kernel reads
 \begin{align}
\widehat{P}_\text{S}(x,\alpha_s) = 
  \begin{pmatrix}
    P_{qq} & 2 N_f P_{qg}
\\
    P_{gq} & P_{gg} 
  \end{pmatrix} \,.
\end{align}
Alternatively, one could consider the  Bjorken-$x$ weighted energy correlation in Mellin space, defined in Eq.~(\ref{eq:mellin-EEC}).
To this end, we first introduce the Mellin moment of NECs,  
\begin{align}
\mathscr{F}^{\text{EC}}_{i/N}\left(N,\ln \frac{E_a \sin\theta}{\mu}\right)
=\int_0^1 \rd z&\, z^{N-1} 
\nn\\
\times&\mathscr{F}^{\text{EC}}_{i/N}\left(z,\ln \frac{E_a \sin\theta}{\mu}\right)\,,
\end{align}
where we reparametrize the NECs  by partonic logarithm 
\begin{align}
\mathscr{F}^{\text{EC}}_{i/N}\left(z,\ln \frac{E_a \sin\theta}{\mu},\mu\right)
=
F^{\text{EC}}_{i/N}\left(z,\ln\frac{E_N \sin \theta} {\mu},\mu\right) 
\,,
\end{align}
with $E_a= z E_N$ as $z$ denotes the momentum fraction of the struck parton.
In DIS, the energy of the active parton is $E_a=Q/2/\hat x$, where $\hat x= x/z$.
The NECs, when parametrized in terms of a partonic logarithm rather than hadronic variables, 
obey a modified DGLAP evolution equation.
Writing $L=\ln(E_a\sin\theta/\mu)$,
their evolution takes the form
  \begin{figure*}[htbp]
  \centering
  \subfloat[]{\includegraphics[width=0.22\textwidth]{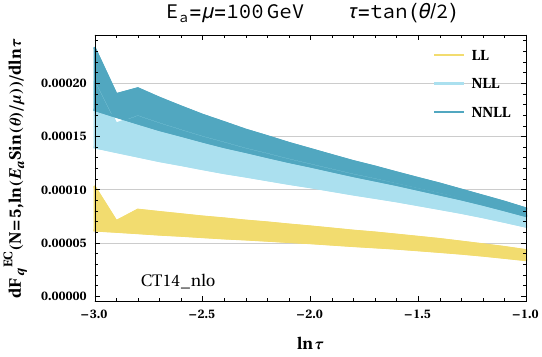}}\hfill
  \subfloat[]{\includegraphics[width=0.22\textwidth]{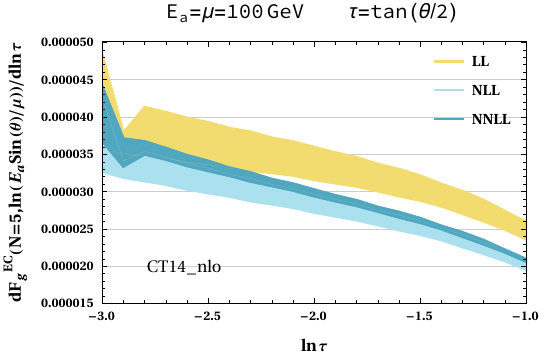}}\hfill
  \subfloat[]{\includegraphics[width=0.22\textwidth]{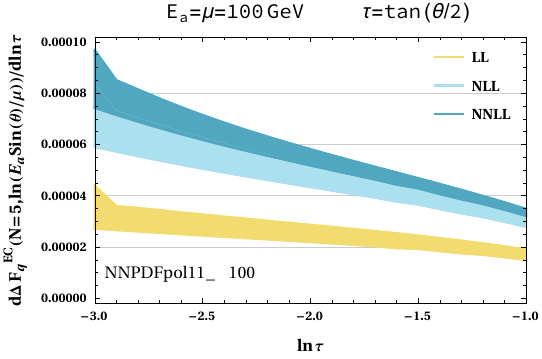}}\hfill
   \subfloat[]{\includegraphics[width=0.22\textwidth]{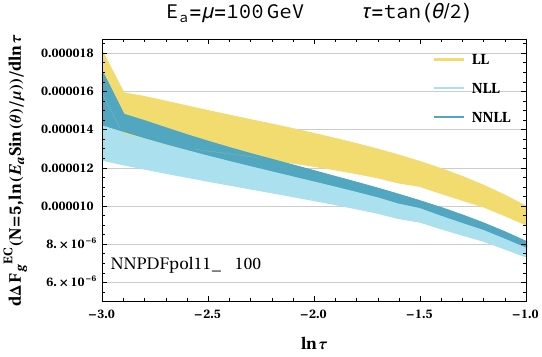}}\hfill
 \caption{Mellin moments of NEC singlets renormalized at a perturbative scale with struck parton energy $E_a=100$GeV.
Panels (a) and (b) show the shape of the unpolarized quark and gluon NECs, $\mathscr{F}^{\text{EC}}_{q,g/N}$, while panels (c) and (d) show the corresponding polarized NECs.
The scale band indicates the variation with respect to the boundary scale $\mu_0 \sim E_a \sin\theta$. 
}
    \label{fig:NEEC_100GeV}
\end{figure*}
\begin{figure}
\centering
\includegraphics[width=0.3 \textwidth]{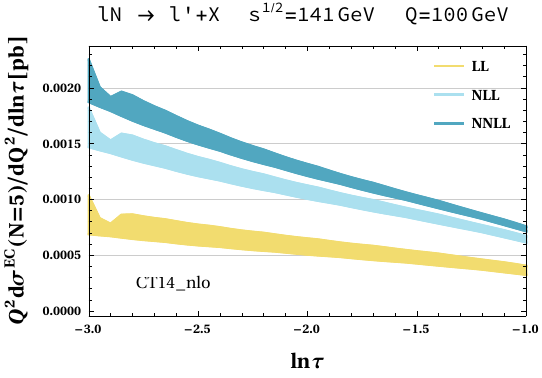}\hspace{0.05\textwidth}
\includegraphics[width=0.3 \textwidth]{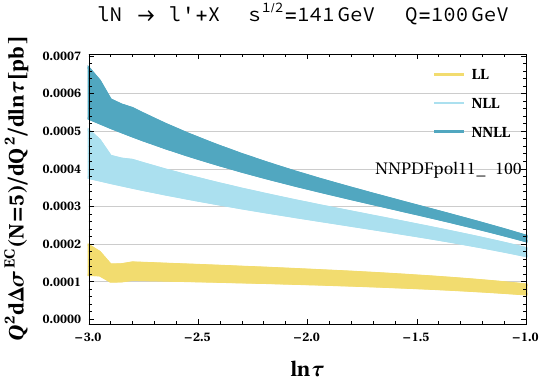}\hspace{0.05\textwidth}
\includegraphics[width=0.3 \textwidth]{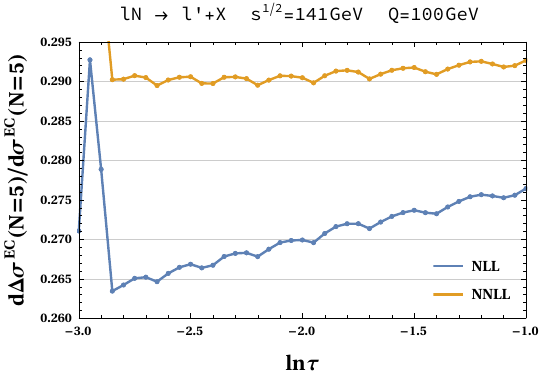}
\caption{Unpolarized and polarized energy correlations to NNLL accuracy in the collinear limit.
The last plot shows the ratio of polarized to unpolarized energy correlations in the forward region. 
The zigzag structure appearing at extremely forward angles reflects the enhanced sensitivity of this region to higher-twist contributions, 
which are not included in the present analysis.
}
\label{fig:EEC-Forward}
\end{figure}
\begin{align}
\label{eq:nec-UV-RG}
&\left.
\frac{\rd}{\rd\ln\mu^2}
\right|_{E_a,\theta}
\mathscr F^{\mathrm{EC}}_{i/N}(N,L,\mu)
\nonumber\\
&\quad =
\sum_j\int_0^1\rd\xi\,\xi^{N-1}
P_{ij}(\xi,\mu)\,
\mathscr F^{\mathrm{EC}}_{j/N}
(N,L-\ln\xi,\mu)
\,,
\end{align}
where $P_{ij}$ are the ordinary spacelike splitting
functions.

The logarithmic translation is  diagonalized by
a Fourier transform,
\begin{equation}
\label{eq:nec-fourier}
\widetilde{\mathscr F}^{\mathrm{EC}}_{i/N}(N,k,\mu)
=
\int_{-\infty}^{\infty}\rd L\,e^{-ikL}
\mathscr F^{\mathrm{EC}}_{i/N}(N,L,\mu)\,.
\end{equation}
In Fourier-$k$ space, the shift $L\to L-\ln\xi$
produces a factor $\xi^{-ik}$, changing the Mellin
weight from $\xi^{N-1}$ to $\xi^{N-ik-1}$.
The evolution therefore probes the ordinary spacelike
splitting matrix at the shifted complex moment $N-ik$.
Collecting the NECs into a column vector, the solution reads
\begin{align}
\label{eq:nec-k-solution}
\widetilde{\boldsymbol{\mathscr F}}^{\mathrm{EC}}_N
(N,k,\mu)
={}&
\left(\frac{\mu}{\mu_0}\right)^{ik}
\mathbf U^{N-ik}(\mu,\mu_0)
\widetilde{\boldsymbol{\mathscr F}}^{\mathrm{EC}}_N
(N,k,\mu_0),
\end{align}
where
\begin{equation}
\label{eq:nec-evolution-operator}
\mathbf U^M(\mu,\mu_0)
=
\mathcal T
\exp\!\left[
\int_{\mu_0}^{\mu}\rd\ln\bar\mu^2\,
\mathbf P(M,\bar\mu)
\right]\,,
\end{equation}
is the ordinary spacelike DGLAP evolution operator, with
$\mathbf P(M,\mu)
=\int_0^1\rd\xi\,\xi^{M-1}\mathbf P(\xi,\mu)$.
The ordering symbol $\mathcal T$ places later evolution
times to the left.
The phase in
Eq.~\eqref{eq:nec-k-solution} accounts for
$\rd L/\rd\ln\mu^2=-1/2$ at fixed $E_a\sin\theta$.

Defining $L_0=\ln(E_a\sin\theta/\mu_0)$, the inverse
Fourier transform gives the compact representation
\begin{equation}
\label{eq:nec-L-operator}
\boldsymbol{\mathscr F}^{\mathrm{EC}}_N(N,L,\mu)
={}
\mathbf U^N(\mu,\mu_0)
\boldsymbol{\mathscr F}^{\mathrm{EC}}_N
\!\left(N,L_0-\overleftarrow{\partial}_N,\mu_0\right)\,.
\end{equation}
Here the arrow denotes derivatives acting only on the
evolution matrix to the left, not on the explicit $N$
dependence of the initial NEC; the operator notation
is understood through the inverse Fourier representation.
It's interesting to note that  the FEC and NEC solutions share the same structure~\cite{Zhu:2025qkx},
with their opposite logarithmic translations encoded
in the shifts $N\pm ik$ in Fourier space and
$L_0\pm\partial_N$ in logarithmic space.

Using the Mellin space NECs, 
we obtain the factorization formula as follows:
\begin{widetext}
\begin{align}
\label{eq:fac-N-space}
\frac{\rd \sigma^\mathrm{EC}(N,\theta)}{\rd Q^2} 
\simeq&
\delta_{\lambda_\ell \lambda_{\ell'}}
 \frac{2\pi\alpha^2}{Q^4}
   \int_0^1 \rd \hat x \,
         \sum_{n=0}^2
                 \hat x^{N-1-n}
 \bigg\{
  \sum_{\Gamma=T,L}
            \sigma^n_\Gamma(Q^2)
  \bigg(
  \sum_i Q_i^2\times \,\mathscr{C}_{i;\text{NS}}^\Gamma(\hat x,Q,\mu) 
   \mathscr{F}^{\text{EC,NS}}_{i/N}\left(N-n,\ln\frac{Q/2 \sin\theta}{\hat x \mu}\right)   
   \nn\\
  +  & \frac{1}{N_f} \sum_{i=1}^{N_f} Q_i^2  \times \, \bigg[
  \mathscr{C}_{q}^\Gamma(\hat x,Q,\mu)   \mathscr{F}^{\text{EC}}_{q/N}\left(N-n,\ln\frac{Q/2 \sin\theta}{\hat x \mu}\right) +
   \mathscr{C}_{g}^\Gamma(\hat x,Q,\mu)   \mathscr{F}^{\text{EC}}_{g/N}\left(N-n,\ln\frac{Q/2 \sin\theta}{\hat x \mu}\right)
 \bigg]
  \bigg)
  \nn\\
  + &\,
   \lambda_{\ell} S_{\parallel} \,
    \Delta\sigma^n(Q^2)
  \bigg(
  \sum_i Q_i^2\times \,\Delta\mathscr{C}_{i;\text{NS}}(\hat x,Q,\mu) 
  \Delta \mathscr{F}^{\text{EC,NS}}_{i/N}\left(N-n,\ln\frac{Q/2 \sin\theta}{\hat x \mu}\right)   
  +
  \frac{1}{N_f} \sum_{i=1}^{N_f} Q_i^2  
   \nn\\
  \times  & \, \bigg[
  \Delta\mathscr{C}_{q}(\hat x,Q,\mu)  \Delta  \mathscr{F}^{\text{EC}}_{q/N}\left(N-n,\ln\frac{Q/2 \sin\theta}{\hat x \mu}\right) +
  \Delta \mathscr{C}_{g}(\hat x,Q,\mu)  \Delta   \mathscr{F}^{\text{EC}}_{g/N}\left(N-n,\ln\frac{Q/2 \sin\theta}{\hat x \mu}\right)
 \bigg]
  \bigg)
 \bigg\}\,.
\end{align}
\end{widetext}
Note that the Born-level form factors in Eq.~(\ref{eq:DIS-form})  are decomposed according to the weight of Bjorken $x$:
$
\sigma_\Gamma (y)=\sum_0^2 \sigma_\Gamma^n(Q^2) x^{-n}
$,
this step is necessary because the variable $y = (Q/\Ecm)^2/x = (Q/\Ecm)^2/(\hat{x} z)$.

Precise predictions based on the factorization formula in Eq.~(\ref{eq:fac-N-space}) require both perturbative and nonperturbative inputs. The perturbative inputs are the DIS coefficient functions computed in Refs.~\cite{Blumlein:2022gpp,Vermaseren:2005qc,Moch:2007rq,Moch:2008fj,Zijlstra:1993sh,Vogt:2008yw,Zijlstra:1992qd,Moch:1999eb}. 
The NECs are intrinsically nonperturbative, but in the perturbative regime $Q\theta\gg\Lambda_\text{QCD}$,
they can be approximated by their twist-2 matching onto collinear PDFs, as given in Eq.~(\ref{eq:NEEC-OPE}). 
The computation of the corresponding matching coefficients $\mathcal{I}_{ij}$ is presented in the next section.
The NNLO DGLAP splitting functions~\cite{Moch:2014sna,Moch:2015usa,Blumlein:2022gpp,Blumlein:2021enk,Blumlein:2021ryt,Zhu:2025gts,Behring:2025avs} are also required to solve the RG evolution~\footnote{For the polarized splitting functions, a minor discrepancy in the color-cubic terms exists between Refs.~\cite{Moch:2015usa} and~\cite{Zhu:2025gts}, with the latter independently confirmed in Ref.~\cite{Behring:2025avs}. This discrepancy originates from the treatment of $\gamma_5$
 in dimensional regularization, as discussed in the next section.}.

We compute the NNLO twist-2 matching of NECs onto collinear PDFs using Eq.~(\ref{eq:NEEC-OPE}). Using this result together with the NNLO DGLAP splitting functions from Refs.~\cite{Moch:2014sna,Moch:2015usa,Blumlein:2022gpp,Blumlein:2021enk,Blumlein:2021ryt,Zhu:2025gts}, we determine the shape of the NECs in the perturbative region $Q\theta\gg\Lambda_\text{QCD}$, as shown in Fig.~\ref{fig:NEEC_100GeV}.
By combining the NNLL-resummed NECs with DIS coefficient functions as in factorization formula Eq.~(\ref{eq:fac-N-space}),
we can predict the  NNLL   energy correction pattern in the forward direction,  as shown in Fig.~\ref{fig:EEC-Forward}.
The scale uncertainties there  are estimated by varying the initial  and factorization scales around $(\mu_0\,, \mu_f) \sim (Q/2 \sin\theta\,,Q)$.
Unlike the CFR, where Sudakov suppression arises due to the recoil of soft gluons, 
the energy correlator in the forward region is inclusive over soft modes, and soft radiation is suppressed in the TFR due to energy weight. 
As a result, the distribution at small angles is not suppressed.
Together with the back-to-back $\text{N}^{3}$LL prediction in Fig.~\ref{fig:EEC-btob}, 
the spin-dependent energy correlator has provided  state-of-the-art precision for probing  
internal structures of the proton across the full range of kinematic regimes,
from the hard radiation region to the current and target fragmentation regions.
\section{Matching with Collinear Factorization}
\label{sec:matching}
The factorization formulas in the CFR~(\ref{eq:SIDIS-fac}) and TFR~(\ref{eq:fac-x-space}) and (\ref{eq:fac-N-space}) do not require a strict scale hierarchy between $q_T \sim Q/2 \sin\theta$ and $\Lambda_\text{QCD}$.
Nevertheless, in the perturbative regime $Q \gg q_T \gg \Lambda_\text{QCD}$, consistency between the CFR/TFR factorization and the standard collinear factorization implies that the TMDs and NECs admit matching onto conventional PDFs.
The $\text{N}^{3}$LO  twist-2 matching of the TMDs is obtained in~\cite{Luo:2019hmp,Luo:2019bmw,Luo:2020epw,Ebert:2020qef,Ebert:2020yqt,Zhu:2025gts}. On the other hand, it was shown in Refs.~\cite{Liu:2022wop,Cao:2023oef}
that the NECs allow OPE onto energy-weighted collinear PDFs as follows
\begin{align}
\label{eq:NEEC-OPE}
\mathscr{F}^{\text{EC}}_{i/N}&\left(z,\ln \frac{E_a \sin\theta}{\mu},\mu\right)
= 
f_i(z,\mu) -\int_z^1
\frac{\rd \xi}{\xi}
\nn\\
\times\,&
\mathcal{I}_{ij}
\left(
\frac{z}{\xi},
\ln \frac{E_a \sin\theta}{\mu},
\alpha_s(\mu)
\right)
\xi
f_j(\xi,\mu)
+\mathcal{O}
\left(\frac{\Lambda_\text{QCD}}{q_T}\right)
\,.
\end{align}
The  matching coefficient $\mathcal{I}_{ij}$ is insensitive to the  hadronic state $N$,  
and thus can be computed from partonic  NECs and renormalized according to Eq.~(\ref{eq:UV-renor}).
After UV renormalization, the NECs are reparametrized  by the partonic logarithm through 
replacement $\ln \frac{E_N}{\mu}\to \ln \frac{E_a}{\mu} -\ln z$ and matched onto energy densities $\xi f_j(\xi,\mu)$.
The renormalization group equation governing the coefficient function follows directly from Eqs.~(\ref{eq:nec-UV-RG}) and (\ref{eq:NEEC-OPE})
\begin{align}
\label{eq:match-rg}
&\frac{\rd \mathcal{I}_{ij}}{\rd \ln \mu^2}
\left(
z\,,
\ln \frac{E_a \sin\theta}{\mu} 
\right)
= 
\int_z^1 
\frac{\rd \xi}{\xi}
\bigg[
 P_{ik}(\xi)
\nn\\
\times \,&
\mathcal{I}_{kj}\left(
\frac{z}{\xi}\,,
\ln \frac{E_a \sin\theta}{\xi\mu} 
\right)
-
\mathcal{I}_{ik}\left(
\frac{z}{\xi}\,,
\ln \frac{E_a \sin\theta}{\mu} 
\right)
\xi P_{kj}(\xi)
\bigg]
\,.
\end{align}
The first term in Eq.~(\ref{eq:match-rg}) originates from RG flow with respect to the  UV cutoff from above,
while the second term arises from RG flow toward the IR.
The solution  up to NNLO reads
\begin{align}
\label{eq:RG-solution}
\mathcal{I}_{ij}&
\left(
z\,,
\ln \frac{E_a \sin\theta}{\mu} 
\right)
=
\delta_{ij}\delta(1-z)
+
a_s(\mu)
\bigg( 
\mathcal{I}^{(1)}_{ij}(z)
\nn\\
+&
L_q\,
\left(
p_{ij}^{(0)}(z)-P_{ij}^{(0)}(z)
\right)
\bigg)
+
a_s^2(\mu)
\bigg(
\mathcal{I}^{(2)}_{ij}(z)
+
L^2_q
\nn\\
\times
&
\bigg[
\frac{
p_{ik}^{(0)}\otimes p_{kj}^{(0)}(z)
}{2}
+
\frac{P_{ik}^{(0)}\otimes P_{kj}^{(0)}(z)}{2}
-
P_{ik}^{(0)}\otimes p_{kj}^{(0)}(z)
\nn\\
-&
\frac{1}{2}\beta_{0}
\bigg(
p_{ij}^{(0)}(z)-
P_{ij}^{(0)}(z)
\bigg)
\bigg]
+L_q
\bigg[
\mathcal{I}^{(1)}_{ik}
\otimes
p_{kj}^{(0)}(z)
\nn\\
-
&
P_{ik}^{(0)}
\otimes
\mathcal{I}^{(1)}_{kj}(z)
-
\beta_0 
\mathcal{I}^{(1)}_{ij}(z)
+
p_{ij}^{(1)}(z)-
P_{ij}^{(1)}(z)
\nn\\
+&
2
\tilde{P}_{ik}^{(0)}
\otimes
\left(p_{kj}^{(0)}-P_{kj}^{(0)}\right)(z)
\bigg]
\bigg)
+\mathcal{O}(a_s^3(\mu))
\,,
\end{align}
where $a_s(\mu)=\alpha_s(\mu)/(4\pi)$ and $L_q=2 \ln \frac{E_a \sin\theta}{\mu}$.
The modified splitting functions are given by
\begin{align}
p_{ij}^{(m)}(z)=z P_{ij}^{(m)}(z)\,,\quad 
\tilde{P}_{ij}^{(0)}(z)=\ln z \,P_{ij}^{(0)}(z)\,.
\end{align} 
The computation of polarized NECs requires careful treatment of the genuinely four-dimensional objects $\gamma_5$ and the Levi-Civita tensor $\epsilon^{\mu\nu\rho\sigma}$, whose definitions must be consistently extended to $D = 4 - 2\epsilon$ dimensions.
However, the anticommutativity of $\gamma_5$ and the cyclicity of the Dirac trace cannot be simultaneously preserved in dimensional regularization. One  approach is to retain the four-dimensional Dirac algebra of $\gamma_5$ by evaluating the trace from a prescribed ``reading point''~\cite{Korner:1991sx,Kreimer:1993bh}, thereby avoiding ambiguities associated with noncyclic traces. 
This method is employed in Refs.~\cite{Matiounine:1998re,Rijken:1997rg} to obtain the polarized DIS coefficient functions
and DGLAP splitting functions in $\overline{\text{MS}}$ scheme. 
 The reading point prescription is  reexamined in the course of computing helicity TMDs~\cite{Zhu:2025gts}, 
 where we reproduce the two-loop scheme transformation factor $Z_{\text{ps}}^{(2)}$, introduced below.
Another approach is to abandon the anticommutativity of $\gamma_5$, and, instead, read in $\gamma_5$ from the effective vertex. 
This leads to the use of the `t Hooft–Veltman-Breitenlohner-Maison (\textsc{HVBM}) scheme~\cite{tHooft:1972tcz,Breitenlohner:1977hr}, or the closely related \textsc{Larin} prescription~\cite{Larin:1991tj,Zijlstra:1992kj,Larin:1993tq}, for the consistent treatment of $\gamma_5$ in dimensional regularization.
In   the \textsc{HVBM} scheme, one proceeds as~\cite{Ravindran:2003gi} 
\begin{enumerate}
\item [1.] Evaluate integrals first\,.
\item [a)] Use multilinear property $\textrm{Tr}[\slashed{ l_1} \slashed{l_2}\dots \gamma_5 ]= l_1^{\mu_1} l_2^{\mu_2}\times\dots \textrm{Tr}[\gamma^{\mu_1} \gamma^{\mu_2}\dots \gamma_5 ]$\,.
\item [b)] Evaluate  tensorlike Feynman integrals and phase space integrals in $D$ dimensions\,.
\item [2.] Begin with the explicit definition of $\gamma_5$\,.
\item [c)] Replace the $\gamma_5$ matrix by
\begin{align}
\gamma_\mu\gamma_5=\frac{i}{6}\e_{\mu\rho\sigma\tau}\gamma^{\rho}\gamma^{\sigma}\gamma^{\tau} \quad \text{or} \quad \gamma_5=\frac{i}{24}\e_{\mu\rho\sigma\tau}\gamma^\mu\gamma^{\rho}\gamma^{\sigma}\gamma^{\tau}\,.
\nn
\end{align}
\item [d)] Compute trace of Dirac matrix  in $D$ dimensions\,.
\item [e)] Contract the Levi-Civita tensors in {four} dimensions\,.
\end{enumerate} 
In computing polarized NECs, we employ a novel $\gamma_5$ scheme--the \textsc{Larin}$^{+}$ 
prescription introduced by the authors of Ref.~\cite{Gutierrez-Reyes:2017glx}.
This particular scheme is well suited when computing TMDs-like objects,  
as  it avoids the explicit appearance of the Levi-Civita tensor $\epsilon^{\mu\nu\rho\sigma}$, 
where one  replaces the $\gamma_5$ matrix by
\begin{align}
\gamma^{+}\gamma_5 \to \frac{i \e_{\perp }^{\alpha\beta}}{2}\gamma_{\alpha}\gamma_{\beta}\,,
\end{align}
and supplement it by the $D$-dimensional relation
\begin{align}
\e_{\perp }^{\alpha_1\beta_1}\e_{\perp }^{\alpha_2\beta_2}=
-g_{\perp }^{\alpha_1\alpha_2}g_{\perp }^{\beta_1 \beta_2}+g_{\perp }^{\alpha_1\beta_2}g_{\perp }^{\beta_1\alpha_2}\,.
\end{align}
As a matter of fact, we have verified that the \textsc{HVBM} 
and \textsc{Larin}$^{+}$  prescriptions yield identical results for helicity TMDs  at N${}^3$LO~\cite{Zhu:2025gts}.
Both the \textsc{HVBM} and \textsc{Larin}$^{+}$ schemes break the anticommutativity of the $\gamma_5$ matrix, which, in turn,
 leads to a violation of the Adler-Bardeen theorem for the nonrenormalization of the axial anomaly beyond one loop~\cite{Adler:1969er}. 
Therefore, renormalization in the prescribed $\gamma_5$ scheme requires additional evanescent counterterms.
To extract the required evanescent counterterms, 
it is necessary to renormalize physical quantities consistently within both Kreimer’s approach and the \textsc{HVBM} or \textsc{Larin}$^{+}$ schemes to the same perturbative order, and define scheme transformations as ratios between them.
This procedure also allows one to fix the  scheme transformation factors,
 whose matrix elements are nontrivial only  in the $q\to q$ entry
 \begin{align}
Z_5=&\,1+\sum_n \left(\frac{\alpha_s}{4 \pi}\right)^n Z_5^{(n)}\,,
\nn\\
(Z_5^{(n)})_{ik}=&\,\delta_{iq}\delta_{kq}
Z_{qq}^{(n)}
=\,
\delta_{iq}\delta_{kq}
\left\{
Z^{(n)}_{\text{ns},+}
+Z^{(n)}_{\text{ps}}
\right\}
\,.
\end{align}
Through  two loops we have  
\begin{equation}
\label{eq:Zqq}
Z^{\text{s}}_{ik}=\delta_{ik}+\delta_{iq}\delta_{kq}
\bigg(
\left(\frac{\alpha_s}{4\pi}\right)
Z^{(1)}_{\text{ns},+}
+
\left(\frac{\alpha_s}{4\pi}\right)^2
\bigg\{
Z^{(2)}_{\text{ns},+}
+Z^{(2)}_{\text{ps}}
\bigg\}
\bigg)
\,.
\end{equation}
An alternative treatment of $\gamma_5$ in dimensional regularization is discussed in~\cite{Chen:2024hlv}.

Owing to the factorization structure,  the polarized NECs  
are expected to be subjected to a  scheme transformation as follows:
\begin{align}
\label{eq:scheme}
\Delta\mathcal{I}^{\overline{\text{MS}}}(\cdot, L_q)=\left[Z_5\otimes\Delta I^\textsc{HVBM}(\cdot, L_P) \right]\odot  z^{-1}_5\,,
\end{align}
where $ I^\textsc{HVBM}(z, L_P)$ is the matching coefficient parametrized by the hadronic logarithm $L_P=2 \ln \frac{E_N \sin\theta}{\mu}$
and $z_5(z)=z\times Z_5(z)$. The modified convolution algebra is defined by
\begin{align}
I(\cdot,L_P)\odot z^{-1}_5\equiv \mathcal {I}(\cdot,L_q)\otimes z^{-1}_5\,,\quad \mathcal {I}(z,L_q)=I(z,L_P)\,.
\end{align}
 Given that $Z_5$ has nontrivial components exclusively in the $q\to q$ channel, Eq.~(\ref{eq:scheme}) implies
 \begin{align}
 \Delta\mathcal{I}_{qq}^{\overline{\text{MS}}}=&\left[Z_5\otimes\Delta I_{qq}^\textsc{HVBM}\right]\odot  z^{-1}_5\,,
 \quad
  \Delta\mathcal{I}_{qg}^{\overline{\text{MS}}}=Z_5\otimes\Delta I_{qg}^\textsc{HVBM}
  \nn\\
   \Delta\mathcal{I}_{gq}^{\overline{\text{MS}}}=&\Delta I_{gq}^\textsc{HVBM}\odot  z^{-1}_5\,,
   \quad
    \Delta\mathcal{I}_{gg}^{\overline{\text{MS}}}=\Delta I_{gg}^\textsc{HVBM}\,.
 \end{align}
After applying the scheme transformations, we confirm that  $ \Delta\mathcal{I}^{\overline{\text{MS}}}$  obeys the RG solution of Eq.~(\ref{eq:RG-solution})
with NLO splitting functions in $\overline{\text{MS}}$ scheme~\cite{Vogelsang:1996im,Vogelsang:1995vh,Zijlstra:1993sh}.
The complete analytic expression for the polarized coefficient functions
are provided as computer-readable files in the Supplemental Material~\cite{ancfiles}, with numerical routines implemented in \textsc{PolyLogsTools}~\cite{Duhr:2019tlz}.
The logarithmic enhancements in the Regge limit $z\to 0$ are collected in Appendix~\ref{sec: small-x}.
There, it is found that the polarized coefficient functions  scale as $z^0$,
in contrast to the unpolarized coefficient functions, 
which exhibit  a $1/z$ scaling behavior in the high-energy limit.
\section{Summary and Outlook}
\label{sec:summary}
\begin{figure}
\centering
\includegraphics[width=0.23 \textwidth]{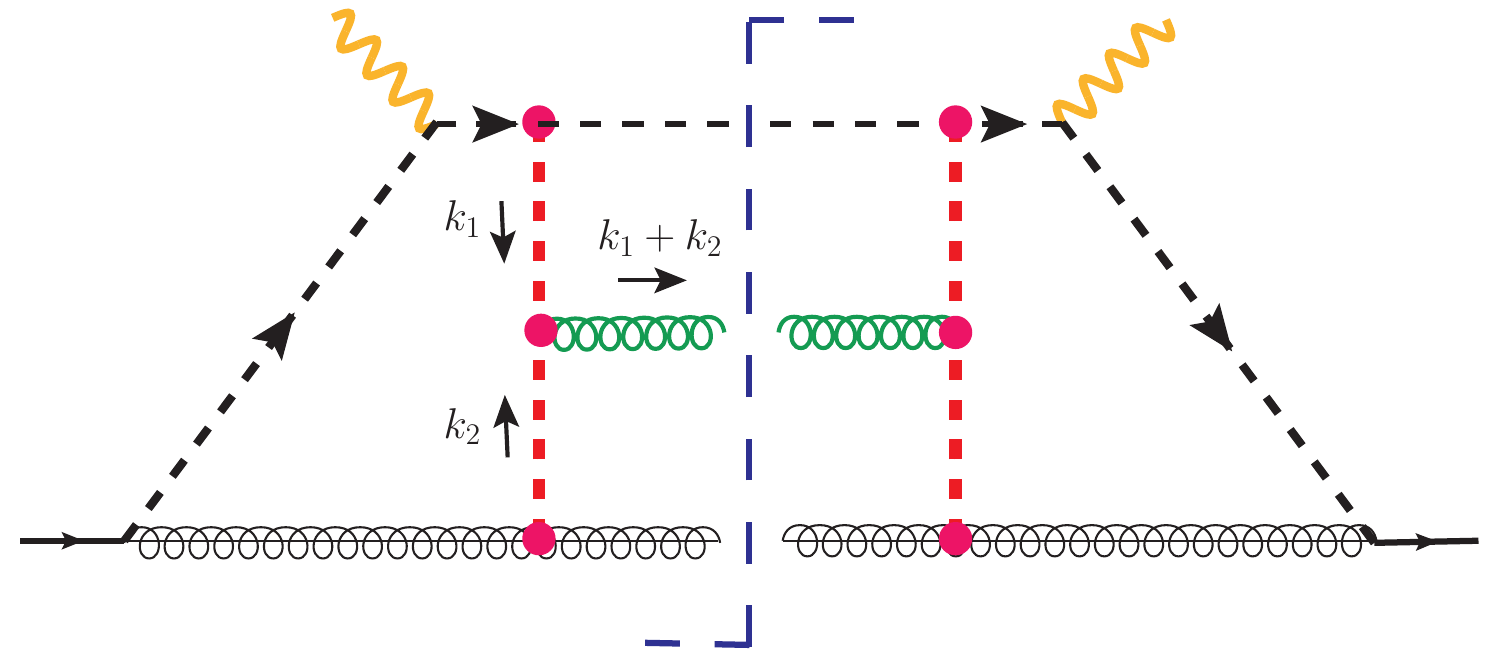}
\hfill
\includegraphics[width=0.23 \textwidth]{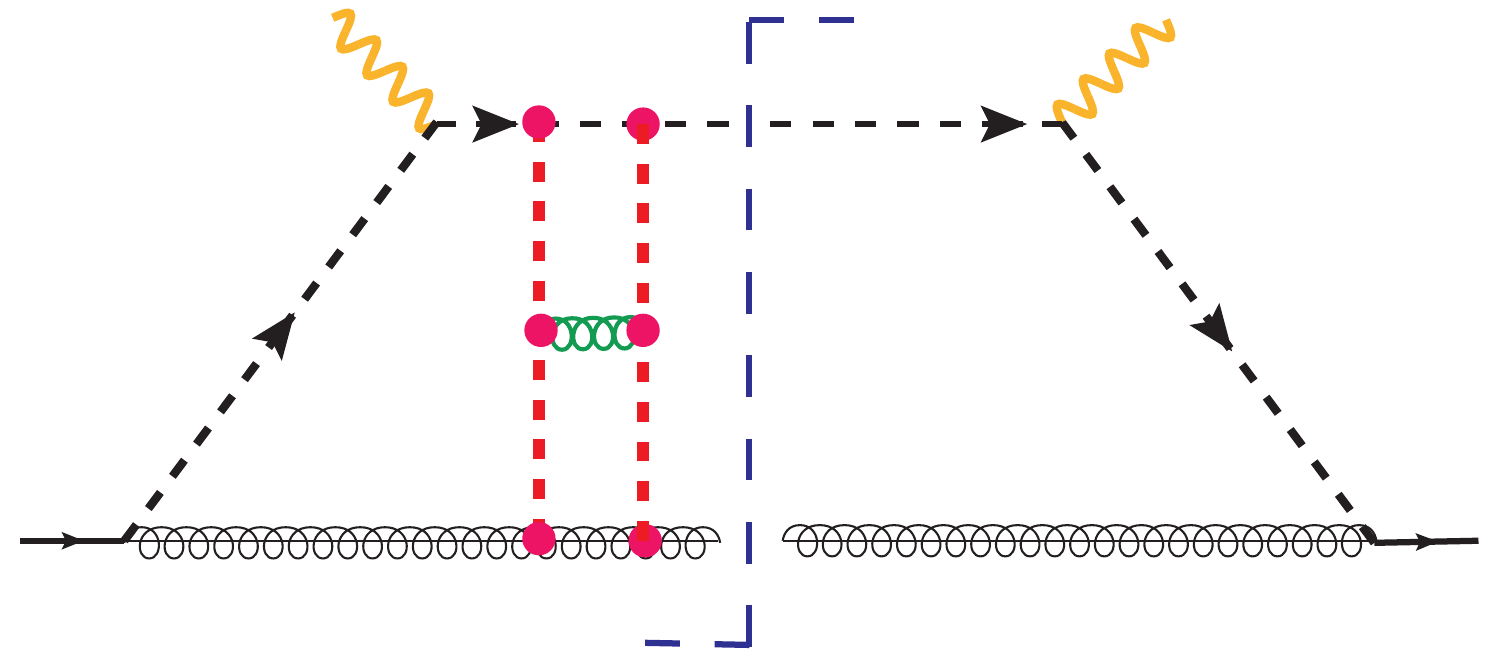}
\caption{Spectator-active interactions through Glauber modes,
with Lipatov vertex off of the Glauber ladders.}
\label{fig:spec-act}
\end{figure}
In this work, we investigate spin-dependent energy correlations in DIS from the current to target fragmentation region.
In the CFR, where the dynamics are governed by TMDs, we showed that energy correlators provide a clean and model-independent probe of TMD physics.
Indeed, by employing  a perturbative $q_T \simeq 3$GeV cutoff,   
we have obtained N${}^3$LL renormalization-group-improved predictions for the wide-angle correlations
without relying on a $b_\ast$ prescription, and we argue
subleading-power corrections in $b_T\Lambda_\text{QCD}$ can be systematically restored by lowering the $q_T$ cut accordingly.
In the TFR, where the dynamics are controlled by NECs, we computed the twist-2 matching at NNLO and achieved NNLL level description for  spectator fragmentation.
Altogether, our study delivers state-of-the-art precision for spin-dependent energy correlation patterns across a broad range of  kinematic regimes, offering an alternative approach to the proton spin decomposition and  proton three-dimensional tomography at the forthcoming EIC,  ideally if one can measure the ECs with arbitrary $N$ and $\theta$.

Looking ahead, several further improvements are worth pursuing. 
First, we assume  Glauber effects drop out for the establishment of factorization.
Although there were substantial arguments~\cite{Collins:1997sr} for this to be correct,
the mechanism responsible for factorization within effective field theories is still missing.
For example, 
Glauber gluons mediate spectator-active interactions,
such as the H diagram in Fig.~\ref{fig:spec-act}.
Depending on the observable, 
their  contributions may be absorbed into collinear Wilson lines~\cite{Rothstein:2016bsq}, 
remain as genuine Glauber contributions~\cite{Gao:2023ivm},
or they cancel out between cut and uncut  diagrams~\cite{Schwartz:2018obd}.
 A  full picture is still unfolding.

Second, 
the unpolarized NECs exhibit  divergences at the Mellin moment $N = 1 \sim x^{-1}\ln x$;
accessing the small-$x$ regime therefore requires an all-order resummation of the corresponding logarithms, 
a program which, in turn, demands a precise definition of the Glauber gluons.
In contrast, the polarized NECs diverge at the Mellin moment $N = 0 \sim x^{0}\ln x$. 
We expect the polarized Regge dynamics to be dominated by subleading fermionic Glauber operators~\cite{Lipatov:2000se,Moult:2017xpp}, 
since the leading $t$-channel Reggeon/Glauber exchanges are insensitive to the proton spin and they cancel in spin asymmetries.
Consequently, contributions that are subleading in the unpolarized case become the leading effects in the polarized small-$x$ regime.
It is also instructive to relate the EFT perspective to the shock-wave formalism with polarized Wilson lines~\cite{Kovchegov:2018znm,Cougoulic:2022gbk,Kovchegov:2015pbl}, which provides a complementary framework for small-$x$ helicity evolution.

Third, the present framework can be extended to T-odd operators, 
thereby enabling dedicated probes of the process dependences of the TMDs and NECs~\cite{Liu:2024kqt}.
For instance, 
the T-odd Boer-Mulders~\cite{Boer:1997nt} and the  Sivers functions~\cite{Sivers:1989cc},
whose existence is attributed to initial- or final-state interactions between the active partons  and the spectators,
are known to exhibit a sign reversal when crossing from SIDIS to the Drell-Yan process~\cite{Collins:2002kn,Brodsky:2002rv,Brodsky:2002cx,Brodsky:2013oya,Belitsky:2002sm}. 
A precise study of these phenomena once again requires a careful definition of the Glauber gluons.

Addressing these points will pave the way toward extending our framework into multi-Regge kinematics and higher-twist domains.
We expect this to substantially broaden the phenomenological reach of energy correlators, particularly for EIC kinematics.
\section*{ACKNOWLEDGMENTS}
We thank Haotian Cao for useful correspondence and for assistance in checking the consistency between the current work and Ref.~\cite{Cao:2023qat}.
The work of J.G. is supported by the National Natural Science Foundation of China (NSFC) under Grant No. 12275173 and the Shanghai Municipal Education Commission under Grant No. 2024AIZD007. H.T.L is supported by the National Science Foundation of China under Grants No.  12275156 and 12321005. 
\appendix 
\section{Form Factors}
\label{sec: form-factor}
To isolate contributions from each of the photon polarization modes, the following bases are chosen for the decomposition of the hadronic tensor
\begin{align}
f_T^{\mu \nu}=&\,\frac{1}{2 x}\left(-g^{\mu\nu}+\frac{q^\mu q^\nu}{q^2}\right)
\nn \\ &
+\left(P_N^\mu+\frac{1}{2x}q^\mu\right)
\left(P_N^\nu+\frac{1}{2x}q^\nu\right)
\frac{1}{P_N\cdot q}\,
\end{align}
\begin{align}
\quad \Delta f^{\mu \nu} =& i \epsilon^{\mu\nu\alpha\beta}\frac{q_\alpha S_\beta}{ x P_N\cdot q} \,,
\nn\\
f_L^{\mu \nu}= &
\left(P_N^\mu+\frac{1}{2x}q^\mu\right)
\left(P_N^\nu+\frac{1}{2x}q^\nu\right)
 \frac{1}{P_N\cdot q}\,.
 \end{align}
\begin{align}
\varepsilon_\mu^0 (\varepsilon_\nu^0)^\ast  f_T^{\mu \nu} =& 0\,,\quad
\varepsilon_\mu^0 (\varepsilon_\nu^0)^\ast f_L^{\mu \nu} =\frac{1}{2 x}\,,\nn \\
\sum \varepsilon_\mu^\pm (\varepsilon_\nu^\pm)^\ast  f_T^{\mu \nu}  =&  \frac{1}{x}\,,\quad
\sum \varepsilon_\mu^\pm (\varepsilon_\nu^\pm)^\ast  f_L^{\mu \nu}  =  0\,.
\end{align}
\begin{align}
L_{\mu\nu} f_T^{\mu \nu}=&\frac{\Ecm^2}{y}\sigma_T\delta_{\lambda_\ell \lambda_{\ell'}}\,,\quad
L_{\mu\nu} f_L^{\mu \nu}=\frac{\Ecm^2}{y}\sigma_L\delta_{\lambda_\ell \lambda_{\ell'}}\,,\nn \\
L_{\mu\nu} \Delta f^{\mu \nu}=&\lambda_{\ell} S_{\parallel} \frac{\Ecm^2}{y}\Delta\sigma\delta_{\lambda_\ell
\lambda_{\ell'}}\,.
\end{align}
Note that $\lambda_{\ell},S_{\parallel}=\pm$ denotes the helicity index of the lepton and proton beams.
\section{Regge limit}
\label{sec: small-x}
The  NECs coefficient functions in Eq.~(\ref{eq:RG-solution}) exhibit logarithmic divergences  in the small-$z$ limit, collected below. 
\begin{widetext}
\begin{align}
\Delta\mathcal{I}^{(1)}_{qq}(z)\simeq
-2C_F(1-2\ln z)\,,
\Delta\mathcal{I}^{(1)}_{qg}(z)\simeq
-4N_f(1+\ln z)\,,
\Delta\mathcal{I}^{(1)}_{gq}(z)\simeq&
4C_F(1+2\ln z)\,,
\Delta\mathcal{I}^{(1)}_{gg}(z)\simeq
8C_A(1+2 \ln z)\,.
\end{align}
\end{widetext}
\begin{widetext}
 \begin{align}
\Delta\mathcal{I}^{(2)}_{qq}(z)\simeq&
\left[
\frac{22}{3}C_A C_F-19 C_F^2
+
\frac{26}{3}C_F N_f
\right]\ln^3 z
+
\left[
\frac{79}{2}C_A C_F
-17 C_F^2+68 C_F N_f
\right]\ln^2 z
\nn\\
+&
\bigg[
C_A C_F \left(
\frac{656}{9}-32\zeta_2\right)
+C_F N_f
\left(
\frac{1375}{9}-16\zeta_2
\right)
+
C_F^2
(-24+24\zeta_2)
\bigg]\ln z
\nn\\
+ &
C_A C_F
\left(
\frac{332}{27}-44\zeta_2-28\zeta_3
\right)
+
C_F^2
\left(
\frac{110}{3}+\frac{100}{3}\zeta_2+12\zeta_3
\right)
+C_F N_f
\left(
\frac{3805}{27}-\frac{16}{3}\zeta_2
\right)
\,.
\end{align}
 \begin{align}
\Delta\mathcal{I}^{(2)}_{qg}(z)\simeq&
\left[
\frac{74}{3}C_A N_f+\frac{13}{3} C_FN_f
\right]\ln^3 z
+
\left[
131C_A N_f-\frac{23}{2} C_F N_f
\right]\ln^2 z
\nn\\
+ &
\bigg[
C_A N_f(276-32\zeta_2)
+C_F N_f
(-72+8\zeta_2)
\bigg]\ln z
\nn\\
+&
C_A N_f
\left(
\frac{716}{3}
-\frac{92}{3}\zeta_2
+
4\zeta_3
\right)
+ 
C_F N_f
\left(
-\frac{239}{3}
+\frac{68}{3}\zeta_2
+12\zeta_3
\right)
\,.
\end{align}
\begin{align}
\Delta\mathcal{I}^{(2)}_{gq}(z)\simeq&
\left[
-\frac{148}{3}C_A C_F-\frac{26}{3}C_F^2
\right]\ln^3 z
+
\left[
-\frac{472}{3}C_A C_F
+6 C_F^2
-\frac{32}{3}C_F N_f
\right]\ln^2 z
\nn\\
+&
\bigg[
-\frac{160}{9}C_F N_f
+
C_F^2(49-48\zeta_2)
+
C_A C_F
\left(
-\frac{3266}{9}+96\zeta_2
\right)
\bigg]\ln z
\nn\\
-&
\frac{272}{27} C_F N_f
+C_F^2
(115-80\zeta_2-64\zeta_3)
+
C_A C_F
\left(
-\frac{10612}{27}
+
100\zeta_2
+32\zeta_3
\right)\,.
\end{align}
 \begin{align}
 \Delta\mathcal{I}^{(2)}_{gg}(z)\simeq&
 \left[
 -\frac{296}{3}C_A^2+\frac{26}{3}C_F N_f
 \right]\ln^3 z
  +
  \left[
  -\frac{701}{3}C_A^2
  -\frac{46}{3}C_A N_f
  +29 C_F N_f
  \right]
  \ln^2 z
  \nn\\
  +&
    \bigg[
    -\frac{142}{3}C_A N_f +C_F N_f (74-16\zeta_2)
    +
    C_A^2
    \left(
    -\frac{1349}{3}
    +64\zeta_2
    \right)
      \bigg]\ln z
      \nn\\
  +&
  C_A^2
  \left(
  -\frac{20285}{54}
  +\frac{20}{3}\zeta_2
  -64\zeta_3
  \right)
  +
  C_A N_f
  \left(
  -\frac{2653}{54}+\frac{40}{3}\zeta_2
  \right)
    +
  C_F N_f
    \left(
    88+8\zeta_2
      \right)\,.
\end{align}
\end{widetext}

\bibliographystyle{apsrev4-1} 
\bibliography{NEEC} 

\allowdisplaybreaks

\end{document}